\documentstyle[preprint,aps]{revtex}

\begin{document}

\draft

\title{Gravitational Radiation from the 
Coalescence of Binary Neutron Stars: Effects Due to 
the Equation of State, Spin, and Mass Ratio}

\author{Xing Zhuge, Joan M. Centrella, and 
Stephen L. W. McMillan}

\address
{Department of Physics and Atmospheric Science,
 Drexel University,
Philadelphia, PA 19104}

\maketitle

\begin{abstract}

We calculate the gravitational radiation produced by the 
coalescence of inspiraling binary neutron stars 
in the Newtonian regime using 3-dimensional
numerical simulations.  The stars are modeled as polytropes 
and start
out in the point-mass regime at wide separation.  The hydrodynamic
integration is performed using smooth particle hydrodynamics (SPH)
with Newtonian gravity, and the gravitational radiation is 
calculated
using the quadrupole approximation.
We have run a number of simulations varying the neutron star
radii, equations of state, spins, and mass ratio.
The resulting gravitational waveforms and spectra are
rich in information about the hydrodynamics of coalescence,
and show characteristic dependence on $GM/Rc^2$, the
equation of state, and the mass ratio.

\end{abstract}


\section{INTRODUCTION}

Coalescing binary neutron stars are among the most promising
 sources
of gravitational waves for detection by interferometers such as 
the Laser Interferometric Gravitational-wave Observatory (LIGO)
\cite{LIGO92}, VIRGO \cite{VIRGO}, and GEO \cite{GEO}.  Recent
 studies
\cite{how-many} suggest that binary inspiral due to
gravitational radiation reaction, and the eventual coalescence 
of the
component stars, may be detectable by these instruments at a rate of
several per year. The inspiral phase comprises the last several
thousand binary orbits and covers the frequency range $f\sim
10$--$1000 {\rm Hz}$, where the broad-band interferometers are most
sensitive.  During this stage, the separation of the stars is much
larger than their radii and the gravitational radiation can be
calculated using post-Newtonian expansions in the
point-mass limit \cite{post-Newt}.  Analysis of the inspiral
wave form is expected to
reveal the masses and spins of the neutron stars, as
well as the orbital parameters of the binary systems
\cite{thorne96a,thorne96b,cutler93,CF94}.

When the binary separation is comparable to the neutron star radius,
hydrodynamic effects become dominant and coalescence takes place
within a few orbits.  The coalescence regime probably lies at or
beyond the upper end of the frequency range accessible to broad-band
detectors, but it may be observed using specially designed narrow
band interferometers \cite{narrow} or resonant detectors
\cite{resonant}.  Such observations may yield valuable information
about neutron star radii, and thereby the nuclear equation of state
\cite{cutler93,KLT94,lindblom92}.

Three-dimensional numerical simulations are needed to study the
detailed hydrodynamical evolution of the coalescence.
Shibata, Nakamura, \& Oohara \cite{SNO,ON} have studied the
behavior of binaries with both synchronously rotating and
non-rotating stars, using an Eulerian code with gravitational
radiation reaction included. Ruffert, et al. \cite{RJS} have
also used Eulerian methods with radiation reaction included to
study coalescence of neutron stars with a physical equation of
state and various spins.
Rasio \& Shapiro \cite{RS92,RS94} have
simulated the coalescence of synchronously rotating neutron-star
binaries using the Lagrangian smooth particle hydrodynamics (SPH)
method with purely Newtonian gravity.   Davies et
al. \cite{MBD93} have carried out SPH simulations of the 
inspiral and
coalescence of nonsynchronously rotating neutron stars, focusing on
the thermodynamics and nuclear physics of the coalescence.
All of these studies used
the quadrupole approximation to calculate the gravitational 
radiation
emitted.  Finally, Wilson, Mathews, and Marronetti \cite{Wilson}
have developed an Eulerian code that incorporates general 
relativistic
effects in the limit in which the metric remains conformally 
flat and
gravitational radiation is neglected.  A multipole expansion 
is used to
calculate the gravitational radiation.

We have carried out 3-D simulations of 
binary neutron star coalescence 
in the Newtonian regime
using SPH, with particular application to
the resulting gravitational wave energy spectrum $dE/df$.  The
neutron stars are initially modeled as spherical polytropes 
 on circular orbits, with separations sufficiently
large that tidal effects are negligible.  The stars thus start out
effectively in the point-mass regime.  The gravitational field is
purely Newtonian, with the gravitational radiation calculated 
using the
quadrupole approximation.  To cause the stars to spiral in, we 
mimic
the effects of gravitational radiation reaction by introducing a
frictional term into the equations of motion to remove orbital 
energy
and angular momentum at the rate given by the equivalent
point-mass inspiral.  As the
neutron stars get closer together the tidal distortions grow and
eventually dominate, and coalescence quickly follows.  The 
resulting
gravitational wave forms match smoothly onto the point-mass 
inspiral
wave forms, facilitating analysis in the frequency domain. 
In Paper I \cite{ZCM94} we considered equal mass
neutron stars with $M$ = 1.4 M$_{\odot}$ and varied the 
neutron star radius
and equation of state. We demonstrated that the resulting 
gravitational wave signatures are rich in information about
the hydrodynamics of coalescence and are sensitive to
both $GM/Rc^2$ and the equation of state.
In this paper, we extend our study
to include the effects of unequal masses as well as spin.

It is important to understand the context of these models.
Our work is carried out in the Newtonian regime and
therefore is a first step toward understanding the
gravitational radiation signatures of binary coalescence.
With these simplified models we are able to study binaries
that start out with fairly wide separations and make
$\sim 3$ orbits before contact, and to concentrate
on the hydrodynamical properties of the merger.
Of course, the Newtonian approximation does break down
for systems involving neutron stars, since 
$GM/Rc^2 \sim 0.2$ for a typical neutron star of
mass $M = 1.4 {\rm M}_{\odot}$ and radius $R = 10 {\rm km}$.
General relativistic effects can therefore be expected to
play an important role in the final stages of inspiral
and coalescence \cite{LW96}, and Newtonian results must be
viewed with appropriate caution.  We believe that 
Newtonian models provide an interesting first look at the
properties of coalescence waveforms and spectra.  In 
addition, they can be used for comparison with general
relativistic calculations to help determine where 
relativistic effects become important and how they
show up in the resulting gravitaional waveforms and
spectra.  Finally, the valuable experience gained in
carrying out these Newtonian calculations is
important for the development of fully general
relativistic models.

This paper is organized as follows. In Sec.~\ref{sim-tech} 
we present
a brief description of the techniques used in our
simulations.  The use of frictional terms in the equations 
of motion
to mimic the effects of gravitational radiation reaction is 
discussed
in Sec.~\ref{inspiral-fric}.  Section~\ref{coal-std} revisits the
standard model (with identical neutron stars having masses
$M = 1.4 {\rm M}_{\odot}$ and radii $R = 10 {\rm km}$),
extending and expanding the analysis begun in Paper I.  
The effects
of changing the neutron star radius, equation of state, and spin
are examined in Sec.~\ref{equal}.  Binaries with unequal mass
components are considered in Sec.~\ref{nonequal}.  Finally,
Sec.~\ref{summary} contains a summary and discussion of our
results.

\section{Simulation Techniques}
\label{sim-tech}

The methods we used to produce our models have been
presented in some detail in Paper I \cite{ZCM94} and reference 
\cite{CM}.  We therefore
give a only brief description of these methods in this section,
and refer the reader to the literature for further information.

Lagrangian techniques such as SPH \cite{SPH} are especially
 attractive
 for modeling neutron star coalescence since the computational
 resources can be concentrated where the mass is located instead 
of
 being spread over a grid that is mostly empty.  We have used the
 implementation of SPH by Hernquist \& Katz \cite{HK} known as
 TREESPH.  In this code, the gravitational 
field is purely Newtonian and
a hierarchical tree method \cite{tree} optimized for vector 
computers
is used to calculate the gravitational forces.   This
leads to a significant gain in efficiency and allows the use of
larger numbers of particles than would be possible with methods
 that
simply sum over all possible pairs of particles.

We calculate the gravitational radiation quantities  in the
quadrupole approximation, which is valid for nearly Newtonian 
sources
\cite{MTW}.  The reduced (i.e., traceless) quadrupole moment
of the source is given by
\begin{equation}
{{I\mkern-6.8mu\raise0.3ex\hbox{-}}}_{ij} = \int\rho \,(x_i x_j -
{\textstyle{\frac{1}{3}}}
\delta_{ij} r^2) \:d^3 r,
\label{Iij}  
\end{equation}
where $i,j=1,2,3$ are spatial indices and $r=(x^2 + y^2 +
z^2)^{1/2}$ is the distance to the source.
For an observer situated on the axis at $\theta=0, \phi=0$ of a 
spherical coordinate system with its origin located at the center
of mass of the source, the gravitational wave amplitudes for
 the two
polarization states are given by
\begin{eqnarray}
h_{+}&=&\frac{G}{c^4}\frac{1}{r}( {\skew6\ddot{
{I\mkern-6.8mu\raise0.3ex\hbox{-}}}}_{xx}- {\skew6\ddot{
{I\mkern-6.8mu\raise0.3ex\hbox{-}}}}_{yy}),
\label{hplus-axis}\\
h_{\times} &=&\frac{G}{c^4}\frac{2}{r} {\skew6\ddot{
{I\mkern-6.8mu\raise0.3ex\hbox{-}}}}_{xy}.\label{hcross-axis}
\end{eqnarray}
Here, an overdot indicates a time derivative $d/dt$.
The standard definition of gravitational-wave luminosity is
\begin{equation}
L = \frac{dE}{dt} = {\frac15}\frac{G}{c^5} \left\langle
 \left\langle
{I\mkern-6.8mu\raise0.3ex\hbox{-}}^{(3)}_{ij}
{I\mkern-6.8mu\raise0.3ex\hbox{-}}^{(3)}_{ij} \right \rangle \right
\rangle ,
\label{lum} 
\end{equation}
where there is an implied sum on $i$ and $j$, 
the superscript $(3)$
indicates the third time derivative, and the 
double angle brackets
indicate an average over several wave periods. 
 Since such averaging
is not well-defined during coalescence, we simply display the
unaveraged quantity $(G/5c^5)
\textstyle{
{I\mkern-6.8mu\raise0.3ex\hbox{-}}^{(3)}_{ij}
{I\mkern-6.8mu\raise0.3ex\hbox{-}}^{(3)}_{ij}}$ in the plots below.
The gravitational wave energy spectrum $dE/df$, which gives the 
energy emitted as gravitational radiation per unit
 frequency interval,
is a key diagnostic tool for understanding the results of our
simulations.  It is given by Thorne \cite{thorne87} in the form
\begin{equation}
\frac{dE}{df} =  \frac{c^3}{G}
 \frac{\pi}{2} (4 \pi r^2) f^2
\langle |\tilde h_+ (f)|^2 + |\tilde h_{\times}(f)|^2 \rangle,
\label{dE/df}  \end{equation}
where $\tilde h(f)$ is the Fourier transform of $h(t)$, and
the angle brackets denote an average over all source angles.
See Paper I for details.

We use the techniques of \cite{CM} to calculate the
 reduced quadrupole moment 
${{I\mkern-6.8mu\raise0.3ex\hbox{-}}}_{ij}$
and its derivatives.  In particular,
${\skew6\dot{I\mkern-6.8mu\raise0.3ex\hbox{-}}}_{ij}$ and
${\skew6\ddot{I\mkern-6.8mu\raise0.3ex\hbox{-}}}_{ij}$ are 
obtained using
particle positions, velocities, and accelerations already 
present in
the code to produce very smooth wave forms.  This yields 
expressions
similar to those of Finn and Evans \cite{FE}.  However,
${I\mkern-6.8mu\raise0.3ex\hbox{-}}^{(3)}_{ij}$ requires the
derivative of the particle accelerations, which 
must be determined numerically and
introduces noise into the gravitational wave
luminosity $L$.  We have applied smoothing to reduce this noise 
in producing all graphs
of $L$ in this paper; see \cite{CM} for further
discussion.

The neutron stars are initially
modeled as widely separated polytropes with equation of state
\begin{equation}
P = K \rho^{\Gamma} = K \rho^{1 + 1/n},
\label{polytrope}  \end{equation}
where $K$ is a constant that measures the specific entropy of the
material and $n$ is the polytropic index.  The stars are placed on
orbits with wide enough
separation that tidal effects are negligible.
An individual star may be allowed to be in uniform rotation about
an axis through its center of mass.  We take the direction of this
spin angular velocity $\Omega_{\rm s}$ 
(which is measured in an inertial frame) to be either parallel or
anti-parallel to the direction of the orbital angular momentum.
 Because the time scale for tidal effects to develop is
far greater than the dynamical time $t_D$ for an individual star,
where
\begin{equation}
t_D = \left ( \frac{R^3}{GM} \right ) ^{1/2} ,
\label{tD}
\end{equation}
we start with stable, ``cold'' polytropes.  
The nonrotating stars ($\Omega_{\rm s} = 0$)
were produced by the method
discussed in \cite{CM}.  The rotating stars were 
produced using the method described in \cite{SHC95} and
 \cite{jlh_phd}.

\section{Modeling Inspiral by Gravitational Radiation Reaction}
\label{inspiral-fric}

Widely separated binary neutron stars (that is, with separation $a
\gg R$) spiral together due to the effects of energy loss by
gravitational radiation reaction.  Once the two stars are close
enough for tidal distortions to be significant, 
hydrodynamical effects
dominate and rapid inspiral and coalescence ensue.  In our
calculations the neutron stars are placed on (nearly)
circular orbits with wide enough separation that 
the stars are effectively in the point-mass limit.
Since the gravitational field is purely Newtonian and does not take
radiation reaction into account, we must explicitly include
these losses to cause inspiral until hydrodynamical effects
take over.

We accomplish this by adding a frictional term to the particle
acceleration equations to remove orbital energy at a rate given by
the point-mass inspiral expression (see \cite{MBD93} for a similar
approach).  The gravitational wave luminosity for
 point-mass inspiral
on circular orbits is \cite{MTW,ST}
\begin{equation}
L_{\rm pm} = \left . \frac{dE}{dt} \right |_{\rm pm} =
\frac{32}{5} \; \frac{G^4}{c^5} \; \frac{\mu^2 {\cal M}^3}{a^5} ,
\label{Lpt}
\end{equation}
where ${\cal M} =
M_1 + M_2$ is the total mass of the system, 
$\mu = M_1 M_2/{\cal M}$ is the reduced mass, 
and the subscript ``pm''
refers to point-mass inspiral. 
We assume that this energy change is due to a frictional force
$\vec f$ that is applied at the center of
 mass of each star, so that
each point in the star feels the same frictional deceleration.
For star 1, we obtain
\begin{equation}
\vec f_1 \cdot \vec V_1 = \left(1 + \frac{M_1}{M_2}\right)^{-1}
\left . \frac{dE}{dt} \right|_{\rm pm} ,
\end{equation}
where $\vec V_1$
is the center of mass velocity of star 1;  an analogous
expression in which the subscripts ``1'' and ``2''
are interchanged holds for star 2.  Since
$\vec f_1$ acts in the direction opposite to
 $\vec V_1$ this gives an
acceleration
\begin{equation}
\vec a_1 = \frac{\vec f_1}{M_1} = -\frac{1}{\cal M}\frac{M_2}{M_1}
\left . \frac{dE}{dt} \right|_{\rm pm} 
\frac{\vec V_1}{|\vec V_1|^2}
\end{equation}
for star 1 and similarly for star 2.
These frictional terms are added to the particle acceleration
 so that all particles
 in a given star experience the same frictional deceleration. 
 The net effect is that the centers of mass of the
stars follow trajectories that approximate point-mass 
inspiral.  The
frictional terms are applied until tidal effects dominate, 
leading to
more rapid inspiral and coalescence by purely Newtonian 
hydrodynamical
processes \cite{LRS3}.
For each of the runs reported in this paper, we
determine the optimal time to turn off the frictional terms 
experimentally; see Paper I for details. 
(Operationally, our assignment of a
particle to a ``star'' is based simply on which body it happened to
belong to initially. Since the frictional term is turned off before
coalescence occurs, the question of what to do after the stars have
merged does not arise.)

The stars are initially placed on the $x$ axis on a 
counter-clockwise circular
orbit with separation $a_0$ in the center of mass frame of
the system in the $x-y$ plane. Thus, the center of mass of $M_1$
is located at $(x,y)$ position $(a_1,0)$ and 
that of $M_2$ is located
at $(-a_2,0)$, where $a_0 = a_1 + a_2$,
$a_1 = a_0 \mu / M_1$ and
similarly for $a_2$. \cite{ST}.
The stars are then given the equivalent point-mass
 circular velocities 
$V_{y,1} = (G/ {\cal M} a_0)^{1/2} M_2 $ 
and
$V_{y,2} = - (G/ {\cal M} a_0)^{1/2} M_1$.

To ensure that the stars start out on the correct 
point-mass inspiral
trajectories, we also give them an initial inward 
radial velocity
$V_x$ as follows.  For point-mass inspiral the 
separation $a(t)$ is
given by \cite{MTW}
\begin{equation}
a(t) = a_0 \left ( 1- \frac{t}{\tau_0} \right )^{1/4} ,
\label{a(t)} \end{equation}
where $a_0$ is the separation at the initial time $t=0$ and
\begin{equation}
\tau_0 = \frac{5}{256} \; \frac{c^5}{G^3} \; 
\frac{a_0^4}{\mu {\cal M}^2}
\label{tau0}  \end{equation}
is the inspiral time, i.e. the time needed to reach separation
$a=0$. We write
\begin{equation}
V_r =\left . \frac{da}{dt} \right |_{t=0} = 
\frac{64}{5} \; \frac{G^3}{c^5} \; \frac{\mu {\cal M}^2}{a_0^3} .
\label{vr} \end{equation}
Requiring the center of mass of the binary system to have zero
velocity then gives $V_{x,1} = -(M_2/{\cal M})V_r$ and
$V_{x,2} = (M_1/{\cal M})V_r$.
The use of the correct initial inspiral 
trajectory allows us to match
our gravitational wave forms smoothly to 
the equivalent point-mass
wave forms.  This is important when analyzing the signals in the
frequency domain.

\section{Binary Coalescence: The Standard Model}
\label{coal-std}

We begin by examining binary coalescence for the case of
equal mass neutron
stars with masses $M_1 = M_2 \equiv M = 1.4 {\rm M}_{\odot}$, radii
$R_1 = R_2 \equiv R = 10 {\rm km}$ (so $GM/Rc^2 = 0.21$),
and polytropic index $n = 1$
($\Gamma = 2$).  
The stars start out with zero spin 
($\Omega_{\rm s} = 0$) on (nearly)
circular orbits with initial separation $a_0 = 40$ km 
in the point-mass limit.
Time is measured in units of the dynamical time $t_{\rm D}$
for a single star using equation~(\ref{tD}); here,
$t_D = 7.3 \times 10^{-5} {\rm s}$.  
We consider this to be our standard run and refer to it as Run 1;
the parameters of this model are
summarized in Table~\ref{models-param}.  This run was first 
discussed in Paper I, which also included tests of our numerical
method with varying particle number and artificial viscosity
coefficients.  In this section we revisit the standard model,
extending and expanding the analysis begun in Paper I,
defining our terminology, and reintroducing some key features
of the problem.

\subsection{General Features of the Coalescence}
\label{gen-features}

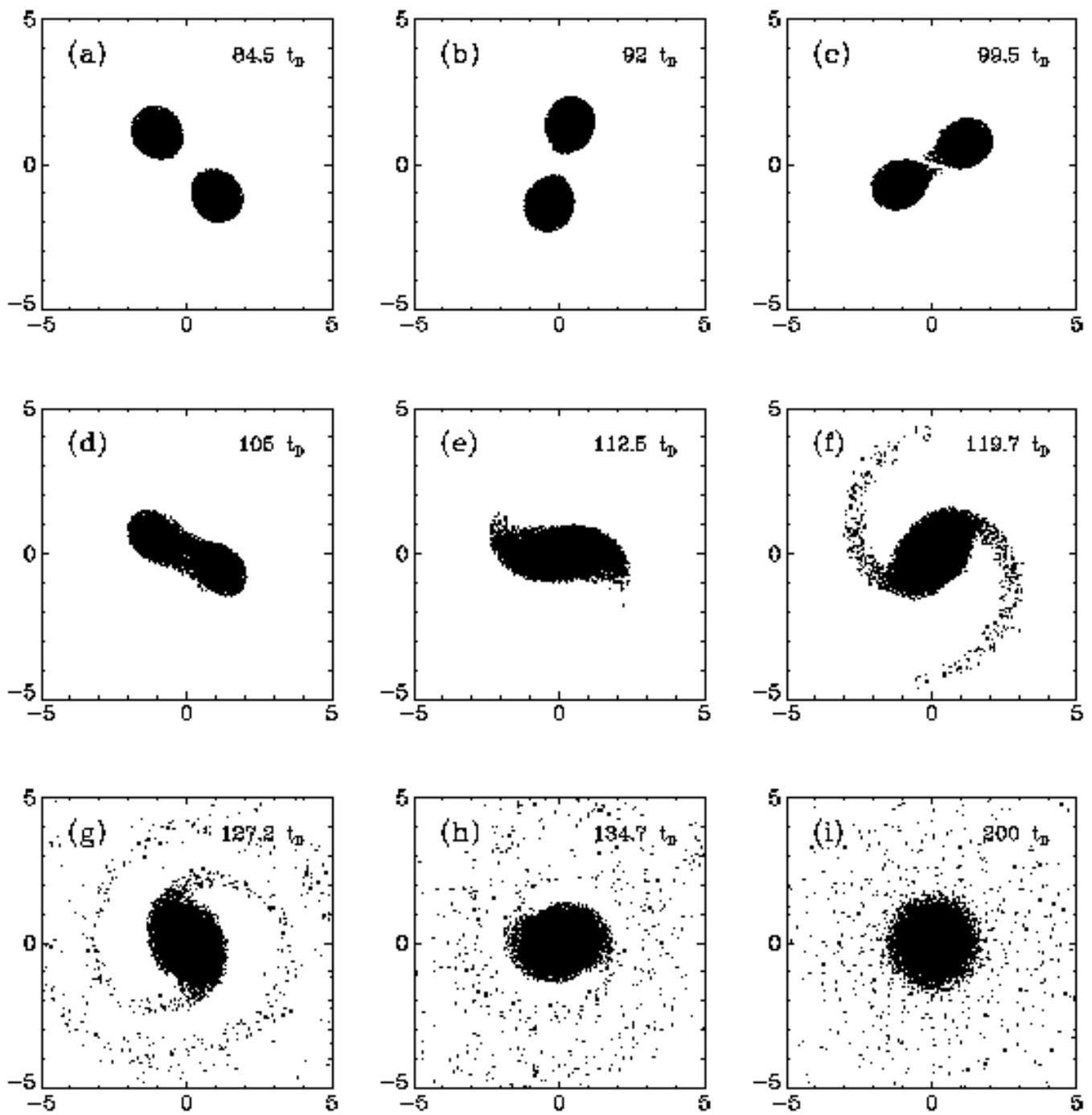
\begin{figure}[p]
\caption{Particle positions are shown projected onto the $x-y$
plane for Run 1.  Here, $M = 1.4 {\rm M}_{\odot}$, $R = 10$ km,
$\Gamma = 2$,
and $t_{\rm D} = 0.073$ ms.  The initial separation $a_0 = 4R$.
The stars are orbiting in the counter-clockwise
direction.  The vertical axis in each frame is $y/R$ and the 
horizontal axis is $x/R$.
\label{run1-snaps}}
\end{figure}

The evolution of this model with $N = 4096$ particles per star is
shown in 
Figure~\ref{run1-snaps}.  In each frame all particles are
projected onto the $x-y$ plane. As the stars spiral 
together, their
tidal bulges grow.  By $t=100 t_{\rm D}$, the center-of-mass
separation of the two stars is $\sim 2.5 R$.  At this
 point the stars
undergo a dynamical instability driven by Newtonian tidal forces
\cite{LRS3}, causing the stars to fall together faster
 than they would
on point-mass orbits.  We therefore turn off the frictional term
in the code at $t=100 t_{\rm D}$ and follow the rest 
of the evolution
using purely Newtonian hydrodynamics and gravity. 
 The stars rapidly
merge and coalesce into a rotating barlike structure.  Spiral arms
form as mass is shed from the ends of the bar.  Angular momentum is 
transported outward by gravitational torques and lost to the spiral
arms.  The arms expand and merge to form a disk around the central
object.  At the end of the run, the system is roughly axisymmetric.

The tidal interactions between the stars increase as they spiral
together. 
Even in the absence of fluid viscosity, Lai and Shapiro 
\cite{LS95} have shown that 
a dynamical tidal
lag angle $\alpha_{\rm dyn}$ develops due to 
the finite time needed for the structure of the stars to
adjust to the rapidly changing tidal potential.
This leads to the formation of tidal bulges that
are not directly aligned, and the resulting gravitational torques
cause each star to spin.  As the stars spiral
together and the separation decreases, $\alpha_{\rm dyn}$ becomes 
larger.  In Figure~\ref{run1-snaps} (c) we estimate the lag angle
of the stars near contact to be $\sim 15^{\circ}$.
For non-spinning stars with the parameters of Run 1,
 Lai and Shapiro
\cite{LS95}
find $\alpha_{\rm dyn} \sim 12^{\circ}$ at contact, 
in good agreement with our results.

Now consider the coalescence in a reference frame co-rotating
with the binary.  As seen from this co-rotating frame, 
the stars appear to be spinning in opposite directions.
As the stars begin to merge, the fluid in star 1 is moving in the
direction opposite to the fluid in star 2 at the point of
contact. This
velocity difference over a short spatial scale 
gives rise to a shear
layer between the stars, which is subject to the Kelvin-Helmholtz
instability \cite{buoyant,vort}.
For real neutron
stars, vortices can form within this layer on small
scales; these eddies can grow and merge together, and turbulence
can develop.  The behavior of this turbulent region 
can be important
in determining the mixing of the material from the individual 
neutron stars to produce a final remnant. For example, a turbulent
viscosity could be generated and thereby the final configuration
may not be irrotational.

However, these processes are difficult to model accurately using
3-D numerical simulations due to their limited
resolution \cite{RS-proc,FR-pvt}. For example,
vortices can form on spatial scales determined
by the resolution of the model and spurious numerical
shear viscosity, rather than on the smaller 
physical scales expected
in real fluids. 
These shortcomings must be taken into account when 
interpreting the
results of numerical models of binary coalescence.

We have examined
our simulations to see if these effects are occurring.
We find that, as the stars merge, a couple of
eddies form
across the shear layer where the two stars meet.  
By $t \sim 127 t_{\rm D}$ (Figure~\ref{run1-snaps} [g]), the
stars have coalesced to the point that there is only a single
density maximum at the center of the remnant.
Each star in
Run 1 has $N = 4096$, giving an effective resolution
of $\sim N^{1/3} = 16$ particles across the diameter of each
star.  The Eulerian models of Ruffert, Janka, and Sch\"{a}fer
\cite{RJS} have somewhat better resolution with the diameter of a
single star covering $\sim 20 - 40$ zones.  Their calculations
show the development of two eddies across the shear layer.
Recent work by Rasio \cite{FR-pvt},
 using SPH with a larger $N$ and 
increased resolution of the shear layer due to the placement of
many smaller mass particles in the outer regions of each star,
shows the development of more eddies on smaller scales.

For these reasons, we suspect that numerical effects 
may be influencing the behavior of the shear layer
in our models. For example, they might cause the stars to mix
more rapidly than they would in reality.  Also, the bar
phase of the evolution might be of shorter duration, and the final 
remnant might have different properties.  
However, Ruffert, et al. \cite{RJS}
point out that, in real neutron stars, the Kelvin-Helmholtz 
mechanism may develop into the macroscopic regime on the spatial 
scale of the coalescing stars as the eddies grow 
and merge.  This could 
produce a final flow pattern similar to that seen in 
the simulations.
They also suggest alternative explanations for the 
development of the
final flow pattern.  Clearly, more work is needed to 
resolve these
issues.  In particular, simulations with many more
 particles (and
more grid zones in the Eulerian case) need to be done.  

The two neutron stars merge to form an object of total mass
$\sim$ 2.8 M$_{\odot}$, with $\sim 94\%$ 
($\sim 2.6$ M$_{\odot}$) of the matter and $\sim 74\%$ 
of the angular momentum in a central core
$\varpi \lesssim 2R$, and the remainder in a disk (see Paper I).
As the
evolution proceeds, the core mass and angular momentum remain
nearly constant, while the angular momentum in the disk is
transported outward due to the effects of gravitational
torques.  By the end of the simulation at
$t = 200 t_{\rm D}$, the central core is axisymmetric with
material on the edge of the remnant at $\varpi \sim 2R$
rotating at $\Omega \sim 0.6 t_D^{-1}$,
 near the critical angular velocity
for breakup.
We believe the reason for this 
axisymmetry can be understood as follows.
Consider an equilibrium sequence of uniformly rotating
axisymmetric polytropes parametrized by 
$\beta = T_{\rm rot}/|W|$, where $T_{\rm rot}$ is the rotational
kinetic energy and $W$ is the gravitational potential energy.
As $\beta$ is increased along such a sequence, a point is
eventually reached at which mass is lost at the equator.
Uniformly rotating polytropes with $n \ge 0.808$ 
($\Gamma \le 2.24$) reach this mass-shedding limit 
before the point
at which ellipsoidal configurations can exist
 \cite{tassoul,LRS-APJS}.
Although the central core in our simulation 
does show differential rotation, we
believe that a similar mechanism is operating here, causing 
the core to be essentially axisymmetric at the end of the run.

Stable, nonrotating neutron stars are believed to
have a maximum mass in the range $\sim 1.4$ M$_{\odot}$ 
to $\sim 2.2$ M$_{\odot}$ \cite{GWM,LS},
depending on the equation of state.  
Rotation can increase this by up to 
$\sim 17\%$, yielding a maximum mass $\lesssim 2.6$ M$_{\odot}$
\cite{FIP,FW} (again depending on the
equation of state), for rotation near breakup speed.
Because the gravitational field in these models is 
purely Newtonian,
the merged remnant in these simulations
cannot collapse to a black hole.  However, since
the final core mass is greater than or comparable to
the maximum allowed neutron star mass, 
general relativity may cause a black hole to form.

The ability of rotation to prevent gravitational collapse to a 
black hole can be estimated by examining the dimensionless
parameter ${\cal A} = cJ/G{\cal M}^2$, 
where ${\cal M}$ refers to the
mass of the entire system.  Piran and Stark \cite{PS} modeled
the gravitational collapse of rigidly rotating polytropes with
$\Gamma = 2$ and $0 \le {\cal A} \le 1.5$ using a fully general 
relativistic 2-D axisymmetric code.  They found that for
${\cal A} < {\cal A}_{\rm crit}$, the collapsing object
formed a black hole, 
whereas for ${\cal A} > {\cal A}_{\rm crit}$ the collapse
was halted by centrifugal
forces leading to a bounce with no black hole formation.  For the
cases they considered, they found 
${\cal A}_{\rm crit} \approx 0.8 - 1.2$.

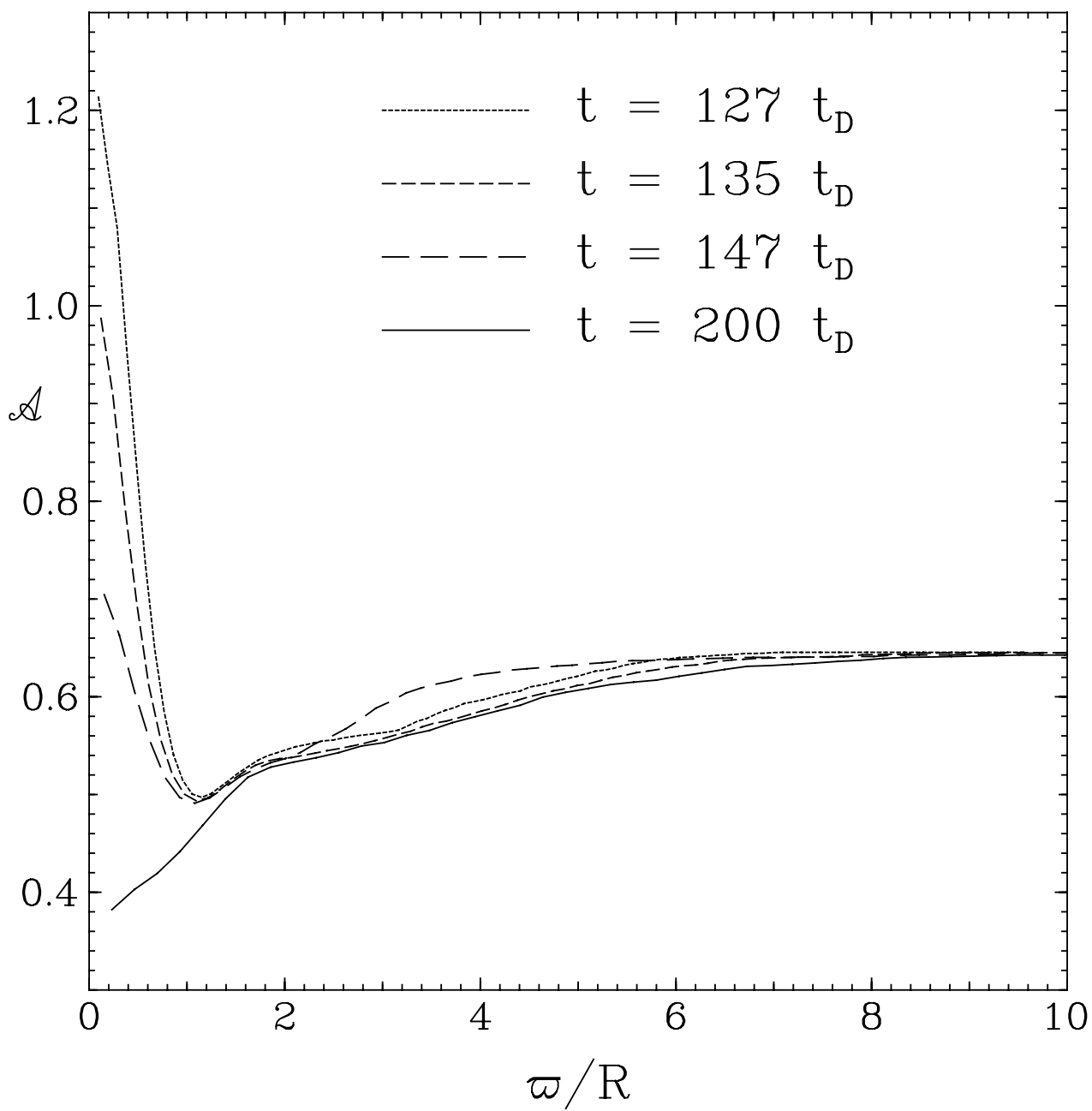
\begin{figure}[p]
\caption{The dimensionless parameter ${\cal A} = cJ/G{\cal M}^2$,
where ${\cal M}$ refers to the mass of the entire system,
is shown as a function of cylindrical radius $\varpi/R$ 
for several times during the coalescence of Run 1.
Here, $R = 10$ km as in Figure~\protect{\ref{run1-snaps}}.
\label{run1-abh}}
\end{figure}

It is interesting to estimate the value of ${\cal A}$ for
our simulations.  Since this rotation parameter is a general
relativistic concept and the values of $J$ and ${\cal M}$
that we determine from our simulations are purely
Newtonian, this discussion must be treated with caution.
With these caveats, we can examine 
Figure~\ref{run1-abh}, which
 shows ${\cal A}$ versus cylindrical radius
$\varpi$ for several times during the coalescence of Run 1.  At 
$t = 127 t_{\rm D}$, the innermost regions have
${\cal A} > 1$. As the evolution proceeds, gravitational torques
transport angular momentum outward and the central 
value of ${\cal A}$
drops.  In all cases, ${\cal A} < {\cal A}_{\rm crit}$ at the core
radius $\varpi \sim 2R$. Thus,
our Newtonian results indicate that it is likely a black hole
will form.  Of course, a firm determination of the final
result of binary neutron star coalescence must await
a fully general relativistic calculation.

\subsection{Properties of the Emitted Gravitational Radiation}

Figure~\ref{run1-gw}(a) shows the gravitational wave 
form $rh_+$ for 
an observer located on the axis at $\theta = \phi = 0$ at distance
$r$ from the source.  (For simplicity, we 
show only one polarization,
$rh_+$. For all runs presented in this paper,
 $rh_{\times}$ is very
similar in appearance to $rh_+$, with a phase
 shift of $90^{\circ}$.)
The solid line gives the code wave form and 
the dashed line the point-mass result.  The code 
wave form matches
the point-mass case for the first couple of orbits 
(note that the orbital period is twice the gravitational wave 
period, $T_{\rm orbit} = 2 T_{\rm GW}$).  As the tidal bulges grow
and the stars spiral together faster than they would on point-mass
trajectories, the gravitational waves increase in both 
amplitude and frequency (cf. \cite{LRS3}).  
Figure~\ref{run1-gw}(b) gives the gravitational wave luminosity
$L/L_0$, where $L_0 = c^5/G$. The code
results (solid lines) initially track the point-mass case (dashed
lines), then depart significantly from the point-mass predictions
somewhat before the onset of dynamical instability.
The amplitudes of the wave forms and luminosity both reach
their maximum values during the early stages of the merger 
at $t \sim 105 - 110 t_{\rm D}$, when the numerical effects 
discussed above are least important.  The amplitudes 
then decrease as the coalescence proceeds.  By 
$t \sim 180 t_{\rm D}$, the gravitational waves have
shut off and the system is essentially axisymmetric.

Table~\ref{models-max} gives the maximum amplitudes of the
gravitational wave forms and luminosities for the runs presented 
in this paper.  
For Run 1, we find that the maximum value
of the wave form for a source located at 
distance r from the observer
is $(c^2/GM)r|h_{\rm max}| \sim 2.0(GM/Rc^2)$.  For
comparison, Rasio and Shapiro (\cite{RS94}; this is listed
as entry RSa in our Table~\ref{models-max}) found
$(c^2/GM)r|h_{\rm max}| \sim 2.4(GM/Rc^2)$ for a synchronous
binary with $\Gamma = 2$.  Also, the maximum luminosity is
$(L_{\rm max}/L_0) \sim 0.39(GM/Rc^2)^5$. 
Rasio and Shapiro found $(L_{\rm max}/L_0) \sim 0.55(GM/Rc^2)^5$.

The gravitational wave energy spectrum $dE/df$ has proved very
useful for analyzing the models in the frequency domain.  For
point-mass inspiral $dE/df \sim f^{-1/3}$ \cite{thorne87},
where the decrease in energy with frequency is due to the fact that
the binary spends fewer cycles (and hence emits less energy) near a
given frequency as it spirals in.  Although our runs do start out in 
the point-mass regime, the stars merge and coalesce within just a 
few orbits.  To achieve a reasonably long
 region of point-mass inspiral
in the frequency domain, we match the code
 wave forms for all runs in this paper onto point-mass
wave forms extending back to much larger binary separations and 
thus lower frequencies.  See Paper I for details.

Figure~\ref{run1-gw}(c) shows the energy 
spectrum $dE/df$ for Run 1.
The short dashed
line is the spectrum for the wave forms truncated at time
$t = 120 t_{\rm D}$; this probes the initial stages of the merger,
during which any numerical effects due to Kelvin-Helmholtz mixing
should be less important.  The solid line shows the spectrum
for the full (i.e. untruncated) wave forms.
In Paper I we defined a number of characteristic
 frequencies based on
features observed in the energy spectrum, and identified these
features with various dynamical phases of the coalescence.
Starting 
in the point-mass regime, as $f$ increases, $dE/df$ first drops
below the point-mass inspiral value and reaches a local minimum
at $f \sim 1500 {\rm Hz}$.  
Recall that, for point masses at separation $a$,
the gravitational wave frequency (which is twice the orbital 
frequency) is given by
\begin{equation}
f_{\rm GW, pm} = \frac{1}{\pi} \;
\left( \frac{G {\cal M}}{a^3} \right)^{1/2}.
\label{f-pm}
\end{equation}
For point-mass
inspiral, the frequency at separation $a \sim 2.5R$ 
(where dynamical
instability is predicted to set in) is 
$f_{\rm dyn} \sim 1500 {\rm Hz}$.
We therefore identify this
dip with the onset of dynamical instability.

Beyond $f_{\rm dyn}$, the spectrum for the truncated wave forms
(short dashed line) rises to a local maximum at $f
\sim 2000$ Hz, then drops off sharply at higher frequencies.
Using equation (\ref{f-pm}), we see that the point-mass
 gravitational
wave frequency at ``contact'' 
(i.e.~at separation $a = 2R$) is $f_{\rm
contact} \sim 2200$Hz.  We associate the peak 
with the formation of a
rapidly rotating merged system.

The solid line in Figure~\ref{run1-gw}(c), 
which shows $dE/df$ for the
entire wave forms, contains features characterizing 
the later stages of
the coalescence.  We see that the peak shifts to
 higher frequencies,
$f_{\rm peak} \sim 2500$Hz, which we attribute 
to the formation of a
transient, rotating barlike structure as the stars
 coalesce, between
$t = 120 t_{\rm D}$ and $t = 150 t_{\rm D}$.  
Continued shrinking of
the merged system as the coalescence proceeds causes 
the rotation
speed of the bar to increase.  We note that, if
 numerical effects due
to Kelvin-Helmholtz mixing are shortening the 
duration of the bar
phase of the evolution, our simulations could be 
substantially underestimating 
the strength of this feature.

Beyond $f_{\rm peak}$, the (untruncated)
spectrum drops sharply, then rises to a
secondary maximum at $f_{\rm sec} \sim 3200$Hz.  
It appears that this
feature is due to transient oscillations induced 
in the merged remnant
during coalescence, and therefore this region of 
the spectrum is undoubtedly
 influenced by the numerical effects
discussed above. 
We point out that, whether or not the final outcome
of actual neutron star coalescence is a black hole,
it seems reasonable that a spectrum of quasinormal
mode oscillations will be produced as the
remnant ``rings down''.
Higher resolution simulations that are
able to track the detailed mixing of the stars during coalescence,
and that fully
incorporate general relativity, are
needed to calculate this region of the spectrum 
accurately.

\section{Binaries with Equal-Mass Components}
\label{equal}

Gravitational radiation from the coalescence of binary
neutron stars is expected to contain important
 information about both the
stars themselves and the interaction between them.
We have parametrized the binary components as polytropes
 of masses $M_1$ and
$M_2$, radii $R_1$ and $R_2$, spins $\Omega_{\rm s,1}$ and 
$\Omega_{\rm s,2}$, and equation of state $\Gamma$.  In this section
we present the results of
runs with components having equal masses
$M_1 = M_2 \equiv M = 1.4 {\rm M}_{\odot}$ and radii
$R_1 = R_2 \equiv R$, and varying $R$, $\Gamma$, 
and $\Omega_{\rm s}$.
The parameters of these runs are summarized in 
Table~\ref{models-param}.

\subsection{Varying the Neutron Star Radius $R$}
\label{vary-R}

Run 2 is the same as Run 1 except that the neutron star radius
$R = 15 {\rm km}$; with $ M = 1.4 {\rm M}_{\odot}$, this gives
$GM/Rc^2 = 0.14$.  Aspects of this run have already
 been presented in Paper I;
we include it here for completeness.  In Run 2, 
the stars start out at initial separation $a_0 = 4R = 60 {\rm km}$
and the gravitational
friction terms are turned off at $t = 245 t_{\rm D}$, 
just before the stars come into contact.
As in the standard run
rapid coalescence takes place, with mass shed
through spiral arms, producing a roughly axisymmetric final
remnant. 

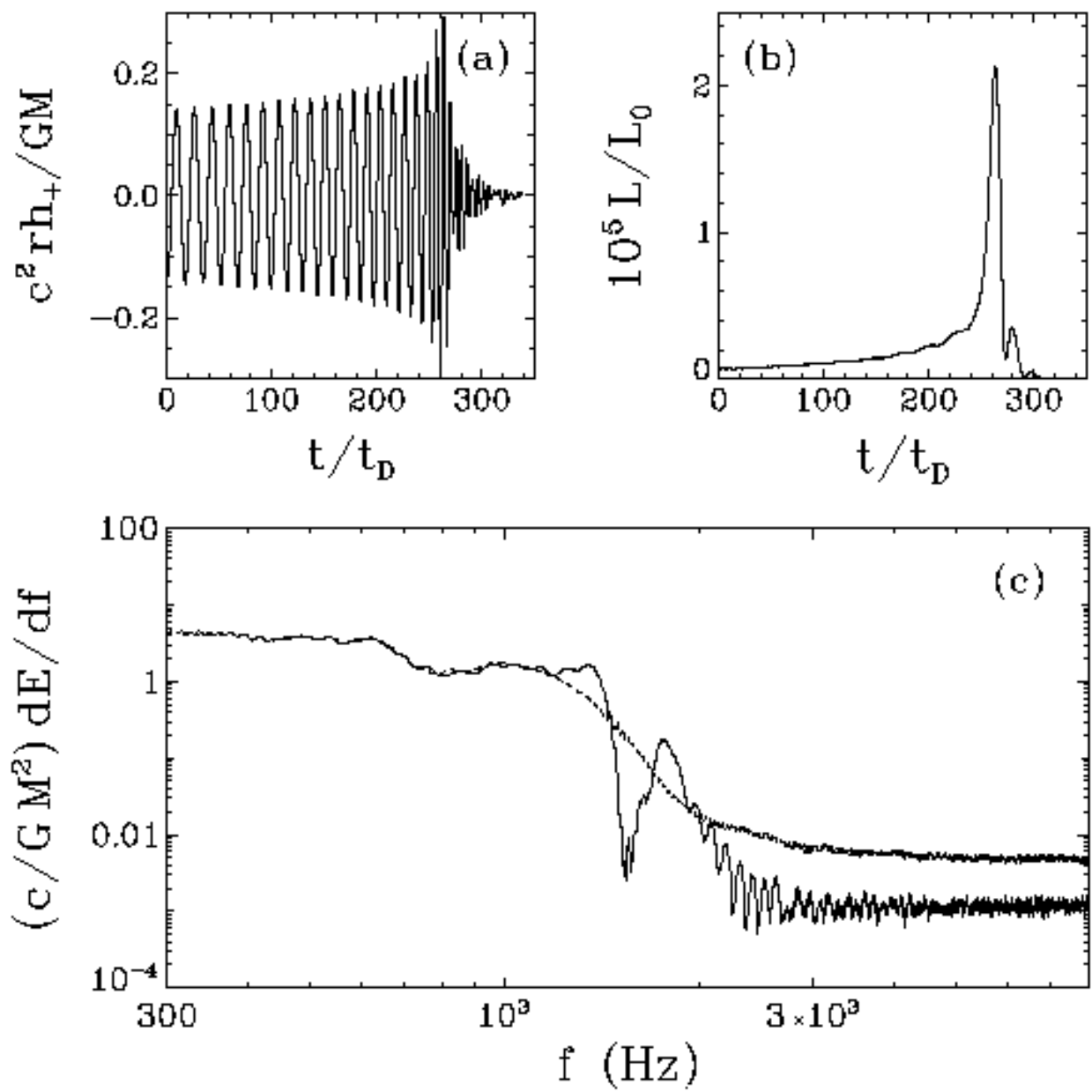
\begin{figure}[p]
\caption{
(a) The gravitational wave form $r h_+$ for Run 2 is shown for
an observer located on the axis at $\theta = \phi = 0$ at distance
$r$ from the source.
(b) The gravitational wave luminosity $L/L_0$,
where $L_0 = c^5/G$.
(c) The gravitational wave energy spectrum $dE/df$. The solid line
shows the spectrum for the entire run, and 
the dashed line shows the
spectrum for the wave forms truncated at $t = 270 t_{\rm D}$.
\label{run2-gw}}
\end{figure}

Figure~\ref{run2-gw}(a) shows
the gravitational wave form $rh_+$ for an observer on the axis at
$\theta = \phi = 0$ at distance $r$ from the source.
The gravitational wave luminosity $L$ is shown 
in Figure~\ref{run2-gw}(b).
 The maximum amplitudes of the wave forms and
luminosity both occur at $t \sim 260 t_{\rm D}$.
The maximum value of the wave form (see Table~\ref{models-max})
for a source located at distance r from the observer
is $(c^2/GM)r|h_{\rm max}| \sim 2.1(GM/Rc^2)$ and
the maximum luminosity is
$(L_{\rm max}/L_0) \sim 0.39(GM/Rc^2)^5$; both of
 these peak amplitudes
are essentially the same as those obtained in the standard run.

The gravitational wave energy spectrum $dE/df$ for Run 2 is shown
in Figure~\ref{run2-gw}(c).  
 For this run, dynamical instability is
expected to occur at separation $a \sim 2.5 R$ \cite{LRS3},
corresponding to a point-mass inspiral frequency $f_{\rm dyn} \sim
850 {\rm Hz}$.  As before, this behavior is evident
in the data as the spectrum
drops below the point-mass value, reaching a minimum at 
$\sim 800-900 {\rm Hz}$.  The truncated spectrum 
(dashed line) then
 rises to a local
maximum at $f \sim 980 {\rm Hz}$ before dropping off
at higher frequencies.  For comparison, the point-mass frequency
at separation $a \sim 2R$ is 
$f_{\rm contact} \sim 1200 {\rm Hz}$.
The solid line in Figure~\ref{run2-gw}(c) shows $dE/df$ for the 
full run; once again, the peak becomes more pronounced and moves to
higher frequencies ($f_{\rm peak} \sim 1350$Hz) when the late-time
wave form is included.
Note that the spectrum does not rise above the point-mass
result
as in Run 1.  Nevertheless, it drops sharply 
just beyond $f_{\rm peak}$,
rising again to a secondary maximum at 
$f_{\rm sec} \sim 1800 {\rm Hz}$. 
The lack of a strong
peak may be due to 
the weaker tidal forces at the point of dynamical instability, 
which occurs at a larger physical separation than in Run 1,
producing a less-pronounced and shorter-lived bar.  As discussed
above, numerical effects may well also be weakening this feature.

It is quite instructive to compare the spectra of Runs 1 and 2.
Since equation~(\ref{f-pm}) gives 
the frequency of gravitational radiation for point-mass inspiral
$\sim a^{-3/2} \sim R^{-3/2}$ at contact, we expect that the 
characteristic frequencies in the Run 2 spectrum should scale
relative to those in Run 1 roughly as $f_2/f_1 \sim 0.54$.
Our results for $f_{\rm peak}$ and $f_{\rm sec}$ do show this
behavior.

\subsection{Varying the Equation of State $\Gamma$}
\label{vary-Gamma}

Run 3 is the same as Run 1, except that we use a stiffer polytropic
equation of state $\Gamma = 3$ ($n = 0.5$).  
A low-resolution version
of this run (with $N = 1024$ particles per star) was 
originally presented in Paper I; here
we repeat this model with
$N = 4096$.

The key difference between the coalescence of equal-mass neutron
stars in Runs 1 and 3 is that the core of the final
 remnant in Run 3
is non-axisymmetric.  This result was also found by Rasio and
Shapiro \cite{RS94} for the case of synchronous binaries with stiff
equations of state.
As remarked above, uniformly rotating polytropes
with $\Gamma > 2.24$ can sustain ellipsoidal shapes, since the
mass-shedding limit along a sequence of equilibrium 
models is reached
after the point at which the ellipsoidal sequence bifurcates
 from the spheroidal sequence.  Although
the coalesced remnant in Run 3 does have some differential rotation,
we expect that this mechanism is operating in this model.  In 
particular, the movement of mass out of the central core region
through spiral arms indicates that the system is rotating at the
mass-shedding limit.  This process is shown in 
Figure~\ref{run1-snaps} (e) - (i) for Run 1.  In Run 3, the
spiral arms are somewhat narrower, as expected for a stiffer
equation of state, and the core of the 
merged remnant is slightly non-axisymmetric.

The signature of this rotating, non-axisymmetric core is clearly
seen in the gravitational wave form $r h_+$, shown in 
Figure~\ref{run3-gw}(a).  After reaching a maximum value at
$t \sim 160 t_{\rm D}$, the amplitude 
initially drops rather sharply, 
and then decays more gradually
on a much longer timescale.
The luminosity $L$ also reaches its maximum value at 
$t \sim 160 t_{\rm D}$, and decreases slowly at late times
(see Figure~\ref{run3-gw}[b]); we find
$(c^2/GM)r|h_{\rm max}| \sim 1.9(GM/Rc^2)$ and 
$(L_{\rm max}/L_0) \sim 0.29(GM/Rc^2)^5$ 
(see Table~\ref{models-max}).

For this run, dynamical instability is expected to occur
at separation $a \sim 2.76 R$ \cite{LRS3}, corresponding to 
a point-mass inspiral frequency $f_{\rm dyn} \sim 1340 
{\rm Hz}$.  Figure~\ref{run3-gw}(c) shows that the spectrum
$dE/df$ for the early stages of the merger
(dashed line) drops below the point-mass
value and reaches a local minimum near $f_{\rm dyn}$.  It then
rises to a local maximum at $f\sim 1800 {\rm Hz}$, 
reaching an amplitude above the point-mass value.  The spectrum 
for the entire run (solid line) has a very pronounced 
peak at $f_{\rm peak} \sim 2200 {\rm Hz}$ due to contributions
 from the rotating, non-axisymmetric core.  

As $\Gamma$ increases the polytrope becomes less centrally
condensed, with a larger fraction of its mass in its outer regions.
Therefore, tidal effects between the two stars become important at
larger separations
 for $\Gamma = 3$ than for $\Gamma = 2$ \cite{Lai-pvt}.
Dynamical instability is thus
predicted to occur at a larger physical separation
in Run 3 than in Run 1, and we expect that 
the spectral features will
appear at correspondingly lower frequencies.  Using point-mass
inspiral we estimate the ratio to be
$f_3/f_1 \sim 0.86$,
and our results do indeed show this behavior.

Run 4 is the same as Run 1, except that $\Gamma = 5/3$
($n = 1.5$) and we used $N = 2048$.  We note in
passing that
the equation of state for nuclear matter is generally believed to 
be stiff, with $\Gamma \gtrsim 2$, and that the case $\Gamma = 5/3$
applied to compact objects
may be more appropriate for low-mass white dwarfs \cite{RS95}.
Nevertheless, this run allows us to explore the effects of a softer 
equation of state on binary neutron star coalescence,
 and we include
it here for comparison.

Rasio and Shapiro \cite{RS95} have shown that, for a
synchronous binary composed of equal-mass
stars with $\Gamma = 5/3$, dynamical instability occurs at
separation $a \sim 2.4R$.  At this time, the stars in 
their model were already in contact. 
In our case, the separation of the centers of mass of the two
stars does drop sharply near $a = 2.4 R$, and the stars are
just on the verge of contact at the time.
As in Run 1, the stars in Run 4 merge to produce a rotating,
axisymmetric final remnant.  During this process, spiral
 arms form and
move mass out of the central region into a flattened halo
 surrounding
the central core.  The spiral arms in Run 4 appear somewhat 
wider 
than in Run 1, as expected, since a softer equation of state
 is used.

The gravitational wave form $rh_+$ 
and the luminosity $L$ are shown in Figure~\ref{run4-gw}(a) and
(b), respectively. Both quantities reach their
maximum values at
$t \sim 115 t_{\rm D}$, with
$(c^2/GM)r|h_{\rm max}| \sim 2.3(GM/Rc^2)$ and
$(L_{\rm max}/L_0) \sim 0.59(GM/Rc^2)^5$.

The spectrum $dE/df$ for Run 4 is shown in 
Figure~\ref{run4-gw}(c). 
Comparison with Figure~\ref{run1-gw}(c) shows that the
spectrum  in Run 4 follows the point-mass
 inspiral result to somewhat
higher frequencies, and therefore smaller separations,
than in Run 1. This is due to the fact that polytropes with
$\Gamma = 5/3$ are more centrally condensed than those
with $\Gamma = 2$, and hence approximate point masses to smaller
separations.
Using the value $a = 2.4 R$ as the separation at 
which dynamical instability takes place, we find
$f_{\rm dyn} \sim 1650 {\rm Hz}$.  The spectrum does dip below
the point-mass result to reach a shallow minimum around
$f_{\rm dyn}$, then rises above it to a broad peak at
$f \sim 2100 - 2500 {\rm Hz}$.  We estimate $f_{\rm peak}$ for the
full spectrum (solid line) to be 
$f_{\rm peak} \sim 2300 {\rm Hz}$.  This is 
followed by a steep drop and a rise to a secondary maximum
at $f_{\rm sec} \sim 4000 {\rm Hz}$.  If we again use the
point-mass scaling $f \sim a^{-3/2}$, along with the above value
of $f_{\rm dyn}$, we find
$f_4/f_1 \sim 1.05$.  Our numerical results
give ratios in the range $\sim 1.05 - 1.20$.

\subsection{Varying the Neutron Star Spin $\Omega_{\rm s}$}
\label{vary-Omega}

Thus far, we have studied neutron binary systems in
 which the components
have zero spin at large separations ($a \gtrsim 4R$), as
seen from a non-rotating frame.  Of course, neutron star binaries
are not expected to be synchronously rotating, since the time
scale for synchronization is in general much longer than the 
time scale for orbital decay and inspiral due to gravitational
radiation reaction \cite{KBC}.  However, it is very likely on both
theoretical and observational grounds that
neutron stars are born with non-zero spin, as evidenced by the many
short-period pulsars known in the Milky Way galaxy.
We have therefore run two models to investigate the effects of
spin on the coalescence and the resulting gravitational wave
signals.

The stars used in these models are taken 
to be uniformly rotating with
spin angular velocity $\Omega_{\rm s}$.  They were 
produced using the method described in \cite{SHC95} (see also
\cite{jlh_phd}), with
$N = 4055$.  (In this method, one does not have complete control
over the number of particles accepted into the star; this accounts
for the somewhat unusual value of $N$.)
We have chosen
$|\Omega_{\rm s}| = 0.175 t_{\rm D}^{-1}$, which is about
$30 \%$ of the maximum rotation rate that a uniformly rotating
neutron star can have before it sheds mass at its 
equator, $|\Omega_{\rm s,max}| \lesssim 0.6 t_{\rm D}^{-1}$
\cite{FIP,CST}.  
For $M = 1.4 {\rm M}_{\odot}$ and $R = 10 {\rm km}$, this
corresponds to a spin period $T_{\rm s} \sim 2.6 {\rm ms}$. 
The stars in our runs have negligible rotational
flattening, with the polar radius $\sim 98\%$ of the equatorial
radius.  Their spins are allowed to be either positive
(i.e.~parallel to the orbital angular momentum) or negative.

In Run 5, both stars are spinning in the positive sense,
with $\Omega_{\rm s,1} = \Omega_{\rm s,2}
= 0.175 t_{\rm D}^{-1}$ at initial separation $a_0 = 4R$. 
 The wave
form $rh_+$ reaches its maximum amplitude at $t \sim 105 t_{\rm D}$
as shown in Figure~\ref{run5-gw}(a).  It attains a higher maximum
amplitude than in Run 1, then drops abruptly to a
 considerably lower
amplitude; cf. Figure~\ref{run1-gw}(a).  The luminosity, which is
presented in Figure~\ref{run5-gw}(b), also reaches a
 larger maximum
value than in Run 1 at $t \sim 105 t_{\rm D}$; see
Table~\ref{models-max}.  We find
$(c^2/GM)r|h_{\rm max}| \sim 2.3(GM/Rc^2)$ and
$(L_{\rm max}/L_0) \sim 0.59(GM/Rc^2)^5$.
The spectrum $dE/df$ is given in Figure~\ref{run5-gw}(c).
Overall, the spectral features are similar in 
appearance to those shown in Figure~\ref{run1-gw}(c) for Run 1. 
However, both $f_{\rm peak}$ and $f_{\rm sec}$ occur at somewhat
higher frequencies in Run 5; see Table~\ref{models-freq}.

Run 6 is the same as Run 5, except that the stars have spins
$\Omega_{\rm s,1} = - \Omega_{\rm s,2} = 0.175 t_{\rm D}^{-1}$.
The gravitational wave form (Figure~\ref{run6-gw}[a]) reaches
its maximum amplitude at $t \sim 100 t_{\rm D}$.  This
maximum value is not as large as that obtained in Run 5, 
and the subsequent
drop in the amplitude of the wave form is more gradual. 
 The luminosity,
shown in Figure~\ref{run6-gw}(b), also reaches a maximum at 
$t \sim 100 t_{\rm D}$.  Here, the maximum luminosity is 
also somewhat
less than in Run 5, although greater than in Run 1, and 
there is no
secondary maximum.  In this case we find
$(c^2/GM)r|h_{\rm max}| \sim 2.0(GM/Rc^2)$,
$(L_{\rm max}/L_0) \sim 0.39(GM/Rc^2)^5$.
Finally, Figure~\ref{run6-gw}(c) shows the spectrum $dE/df$.
 Comparison with Figure~\ref{run1-gw}(c)
shows that the spectral features for Run 6 are again
similar in appearance to those of Run 1 and occur at about the
same frequencies.  These results are summarized in 
Table~\ref{models-freq}.

\section{Binaries with Unequal-Mass Components}
\label{nonequal}

In this section we present the results of simulations with  binary
components of unequal masses.  The primary is taken to have mass
$M_1 = 1.4 {\rm M}_{\odot}$ and radius
$R_1 = 10 {\rm km}$.  We then let the secondary have mass
$M_2 = q M_1$, where $q = M_2/M_1 < 1$ is the mass ratio. 
 To calculate
the radius of the secondary 
$R_2$, we follow Rasio and Shapiro \cite{RS94} and assume
that the system has constant specific entropy.  Thus, the initial
state of both components is constructed using 
equation~(\ref{polytrope}) with the same polytropic constant
$K_1 = K_2$ and the same value of $\Gamma$.  This gives the
mass-radius relation
\begin{equation}
\frac{R_1}{R_2} = \left( \frac{M_1}{M_2} 
\right) ^{(\Gamma - 2)/(3\Gamma - 4)} .
\label{mass-radius}
\end{equation}
Both components are taken to have zero spin, 
$\Omega_{\rm s,1} = \Omega_{\rm s,2} = 0$.  We use $N=4096$,
with all particles in a given star having
the same mass.  The dynamical time $t_{\rm D}$ used to 
describe the
evolution of these runs is calculated using the parameters of the
primary.  Table~\ref{models-param} summarizes the
 parameters of these
models.

For this paper, we have carried out runs with two different
mass ratios, $q = 0.85$ and $q = 0.5$, using both $\Gamma = 2$
and $\Gamma = 3$.  The value $q = 0.85$ is believed to be the most
probable mass ratio for the binary pulsar
 PSR 2303+46 \cite{Thorsett}.
Although this  is the smallest observed value of $q$ for a 
binary pulsar with a
neutron star companion \cite{RS94}, we
consider it important at this early stage in our understanding of
the gravitational wave signals from binary coalescence to explore
more extreme mass ratios.  For this reason we have also
 considered the 
case $q = 0.5$. 

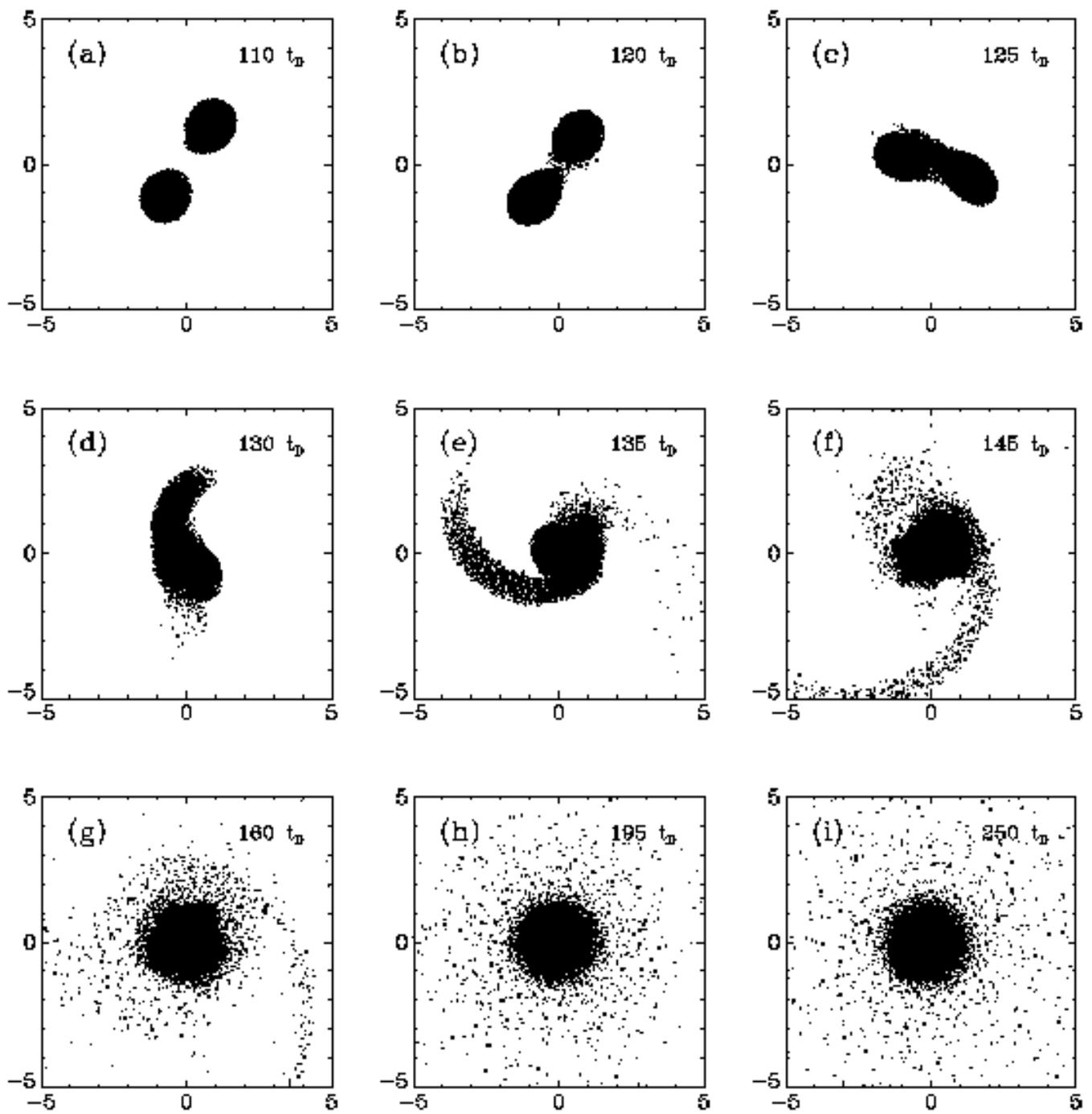
\begin{figure}[p]
\caption{Particle positions are shown projected onto the $x-y$
plane for Run 7.  Here, $M_1 = 1.4 {\rm M}_{\odot}$, 
$M_2 = 0.85 M_1$, $R_1 = R_2 = 10$ km,  $\Gamma = 2$,
and $t_{\rm D} = 0.073$ ms.  The initial separation $a_0 = 4 R_1$.
The stars are orbiting in the counter-clockwise
direction.  The vertical axis in each frame is $y/R_1$ and the 
horizontal axis is $x/R_1$.
\label{run7-snaps}}
\end{figure}

Run 7 is a slightly asymmetric binary with $q = 0.85$ and $\Gamma
= 2$.  By equation~(\ref{mass-radius}), both components have equal
radii, $R_2 = R_1$.  In Figure~\ref{run7-snaps} all particles are
projected onto the $x-y$ plane to show the evolution of this model.
Tidal bulges develop as the stars spiral together, 
and mass transfer 
begins by $t \sim
120 t_{\rm D}$.  As the merger proceeds, a single spiral
arm or elongated ``tail''
is formed from the secondary, seen in Figure~\ref{run7-snaps} (d) 
- (g).  By the end of the simulation at $t = 250 t_{\rm D}$,
the secondary has completely merged with the
primary. The rotating, axisymmetric remnant has a central core of
radius $\sim 1.5 R_1$ 
and is surrounded by a flattened halo consisting of
material from the secondary.

The gravitational wave form $r h_+$ for Run 7 is shown in 
Figure~\ref{run7-gw}(a), and the luminosity $L$ is shown in
Figure~\ref{run7-gw}(b).  Both quantities reach their maximum
amplitudes at $t \sim 125 t_{\rm D}$, during the early stages
of the merger.   
The maximum value of the wave form (see Table~\ref{models-max})
for a source located at distance r from the observer
is $(c^2/GM)r|h_{\rm max}| \sim 1.6(GM_1/R_1c^2)$ and
the maximum luminosity is
$(L_{\rm max}/L_0) \sim 0.18(GM_1/R_1c^2)^5$.
Figure~\ref{run7-gw}(c) shows the gravitational wave energy spectrum
$dE/df$ for this run.

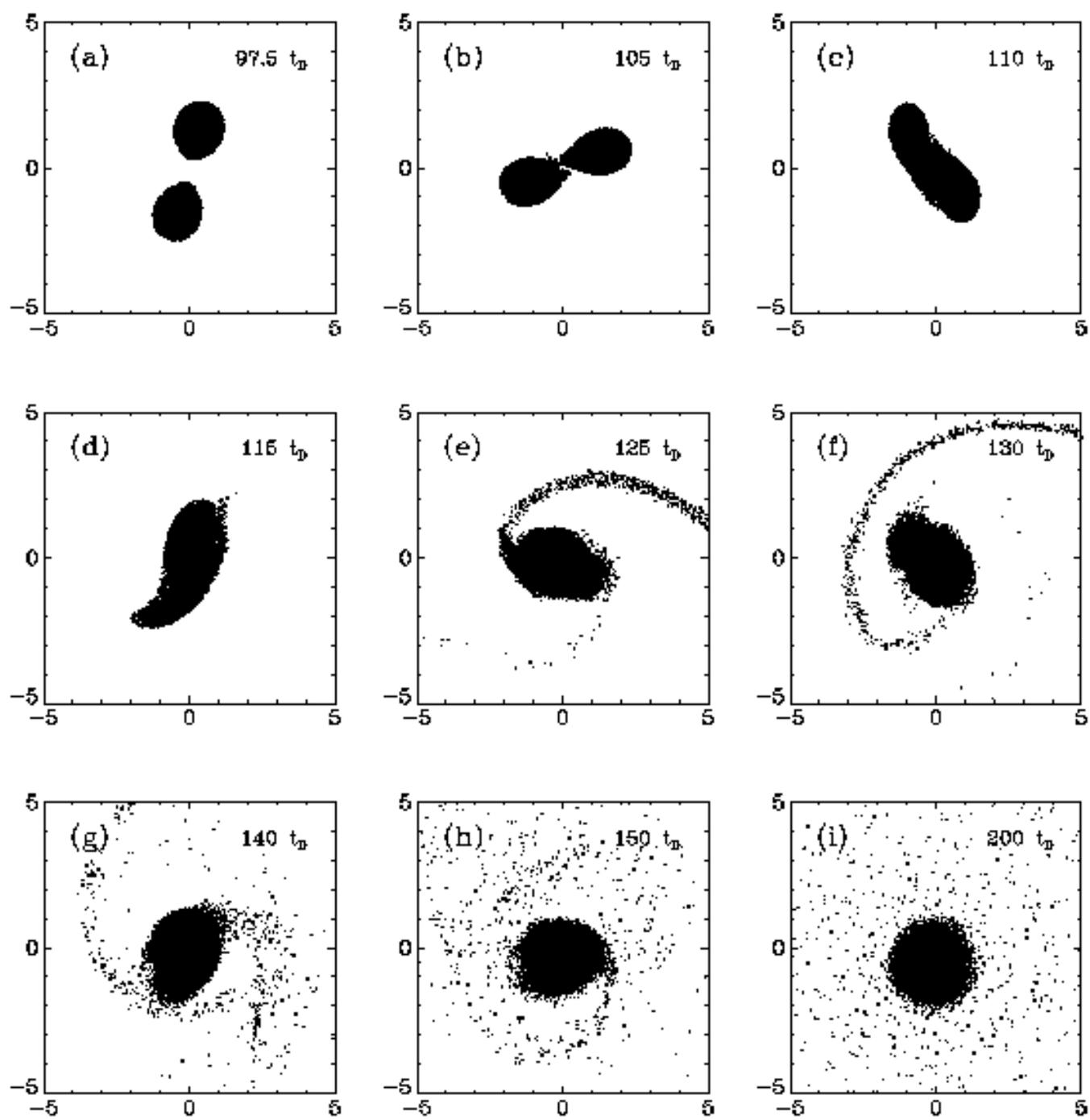
\begin{figure}[p]
\caption{Particle positions are shown projected onto the $x-y$
plane for Run 8.  Here, $M_1 = 1.4 {\rm M}_{\odot}$, 
$M_2 = 0.85 M_1$, $R_1 = 10$ km, $R_2 = 9.7$ km, $\Gamma = 3$,
and $t_{\rm D} = 0.073$ ms.  The initial separation $a_0 = 4 R_1$.
The stars are orbiting in the counter-clockwise
direction.  The vertical axis in each frame is $y/R_1$ and the 
horizontal axis is $x/R_1$.
\label{run8-snaps}}
\end{figure}

In Run 8 $q = 0.85$ and $\Gamma = 3$ so that, according to
equation~(\ref{mass-radius}), $R_2 = 9.7 {\rm km}$. 
Figure~\ref{run8-snaps} shows the evolution of this run, which
proceeds in a similar way to that in Run 7.  
In this case, however,
the stiffer equation of state leads to a narrower extended tail.
The central region also retains an elongated shape for a longer
time.  By the end of the simulation at $t = 200 t_{\rm D}$,
the central core is essentially axisymmetric with radius
$\sim 1.5 R_1$ and a flattened halo composed of mass from the
secondary.  The remnant is also slightly displaced from the sytem
center of mass.  Rasio and Shapiro \cite{RS94} also notice this
behavior in their synchronous merger with $q = 0.85$ and 
$\Gamma = 3$, attributing it to the asymmetric, single-arm
mass outflow.  Interestingly, Run 7 with $\Gamma = 2$ also has
asymmetric outflow, yet the final remnant suffers a much smaller
displacement from the system center of mass.

The gravitational wave form $r h_+$ for Run 8 is shown in 
Figure~\ref{run8-gw}(a), and the luminosity $L$ is shown in
Figure~\ref{run8-gw}(b).  Both quantities reach their maximum
amplitudes at $t \sim 110 t_{\rm D}$, during the early stages
of the merger.  More gravitational radiation is produced after
the maximum is reached than in Run 7, due to the more strongly
non-axisymmetric central region.
The peak value of the wave form (see Table~\ref{models-max})
for a source located at distance r from the observer
is $(c^2/GM)r|h_{\rm max}| \sim 1.6(GM_1/R_1c^2)$;
the maximum luminosity is
$(L_{\rm max}/L_0) \sim 0.19(GM_1/R_1c^2)^5$.
Figure~\ref{run8-gw}(c) shows the gravitational wave energy spectrum
$dE/df$ for this run. Comparing the spectrum for the full 
wave forms with that
given in Figure~\ref{run7-gw}(c) for Run 7, we see that 
Run 8 with $\Gamma = 3$ shows a more pronounced peak at a 
slightly lower frequency $f_{\rm peak}$.
This behavior is similar to that seen
in comparing Runs 1 and 3 and is attributed to the
stronger barlike central region;  cf. \S~\ref{vary-Gamma}.

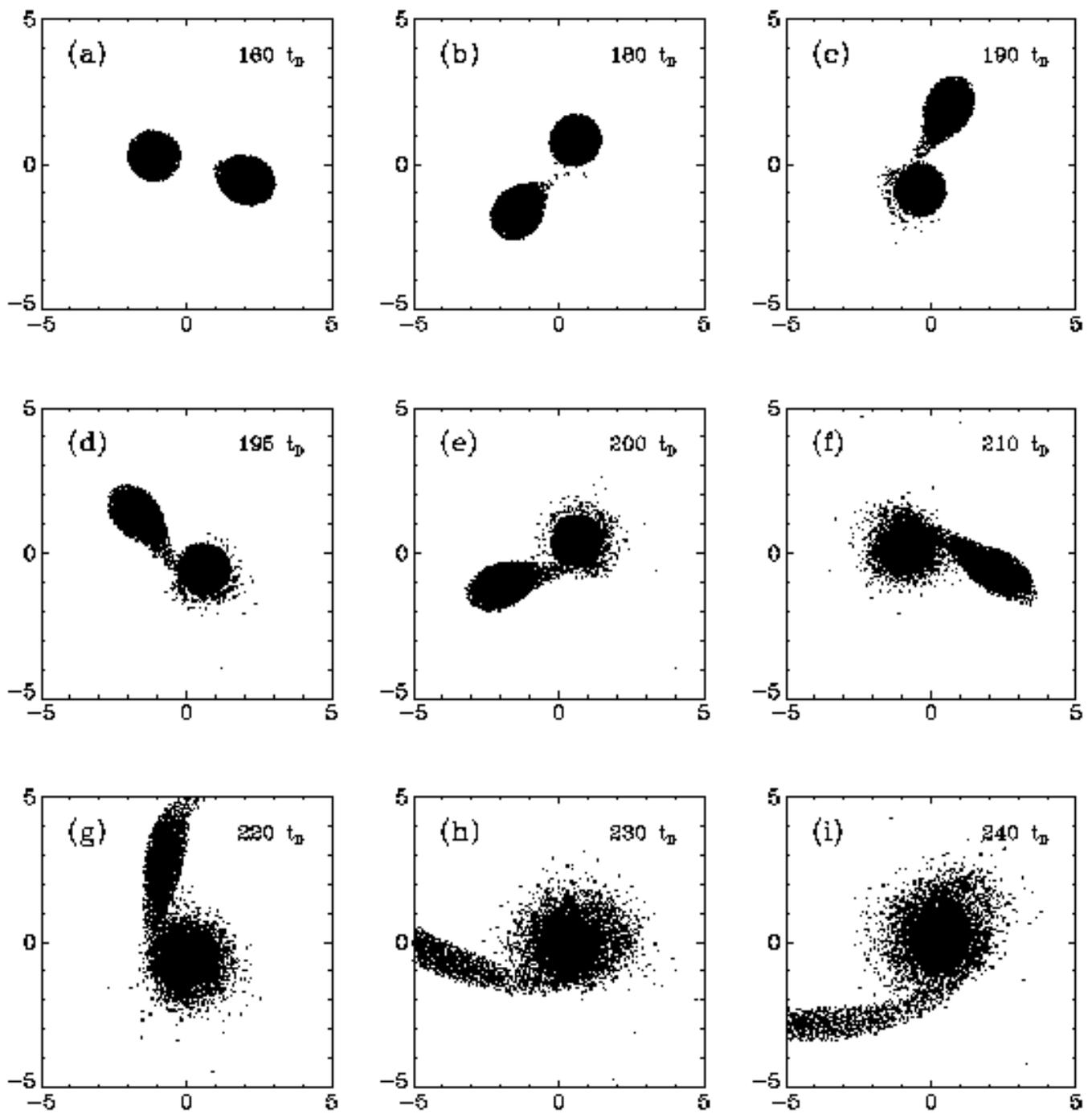
\begin{figure}[p]
\caption{Particle positions are shown projected onto the $x-y$
plane for Run 9.  Here, $M_1 = 1.4 {\rm M}_{\odot}$, 
$M_2 = 0.5 M_1$, $R_1 = R_2 = 10$ km,  $\Gamma = 2$,
and $t_{\rm D} = 0.073$ ms.  The initial separation $a_0 = 4 R_1$.
The stars are orbiting in the counter-clockwise
direction.  The vertical axis in each frame is $y/R_1$ and the 
horizontal axis is $x/R_1$.
\label{run9-snaps}}
\end{figure}

Run 9 has $q = 0.5$ and 
$\Gamma = 2$ which implies
that the components have equal radii, $R_2 = R_1$.  
The evolution of 
this model is shown in Figure~\ref{run9-snaps}, where
 all particles
are projected onto the $x-y$ plane.  As the stars spiral 
together, the
secondary develops a tidal bulge and starts to lose mass to the 
primary.   Although this mass transfer completely disrupts the
secondary, the primary is not strongly affected.  At the
 end of the
run $t = 300 t_{\rm D}$, most of the mass of $M_2$ is spread in a 
flattened halo of radius $\sim 5 R_1$
around $M_1$, with a single spiral arm extending out to
very large radii, up to $\sim 40 R_1$.  The halo and core contain
$\sim 96\%$ of the total system mass, and $\sim 70\%$ of the
total angular momentum (relative to the
center of mass of the system). 

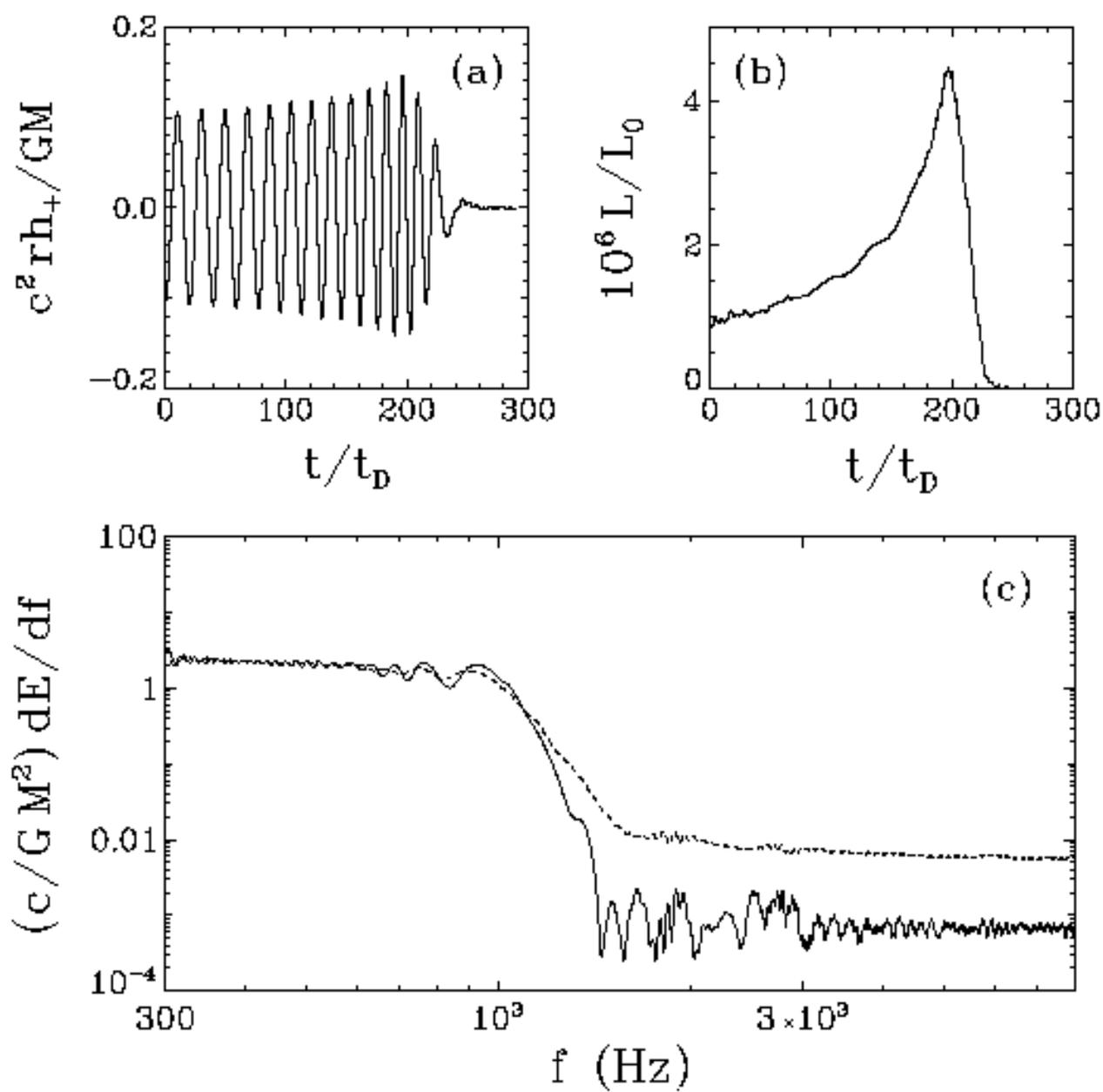
\begin{figure}[p]
\caption{
(a) The gravitational wave form $r h_+$ for Run 9 is shown for
an observer located on the axis at $\theta = \phi = 0$ at distance
$r$ from the source.
(b) The gravitational wave luminosity $L/L_0$,
where $L_0 = c^5/G$.
(c) The gravitational wave energy spectrum $dE/df$. The solid line
shows the spectrum for the entire run, and the 
dashed line shows the
spectrum for the wave forms truncated at $t = 215 t_{\rm D}$.
\label{run9-gw}}
\end{figure}

Figure~\ref{run9-gw}(a) shows the gravitational
 wave form $r h_+$ for 
this run, and Figure~\ref{run9-gw}(b) shows the 
luminosity $L$. Both
the wave form and the luminosity reach their peak values
at $t \sim 195 t_{\rm D}$, during the early stages 
of mass transfer.
The gravitational waves shut off less than 1.5 orbits
 later, indicating
that the secondary is quickly disrupted.
The maximum value of the wave form (see Table~\ref{models-max})
for a source located at distance r from the observer
is $(c^2/GM)r|h_{\rm max}| \sim 0.67(GM_1/R_1c^2)$ and
the maximum luminosity is
$(L_{\rm max}/L_0) \sim 0.011(GM_1/R_1c^2)^5$.
The gravitational wave energy spectrum $dE/df$ for this
 run is shown in
Figure~\ref{run9-gw}(c).  Since the evolution of the 
system proceeds
rapidly by mass transfer, 
there is little difference between these
two curves, with
$f_{\rm peak} \sim 900 {\rm Hz}$; there is no discernible
high-frequency secondary feature present at late times.

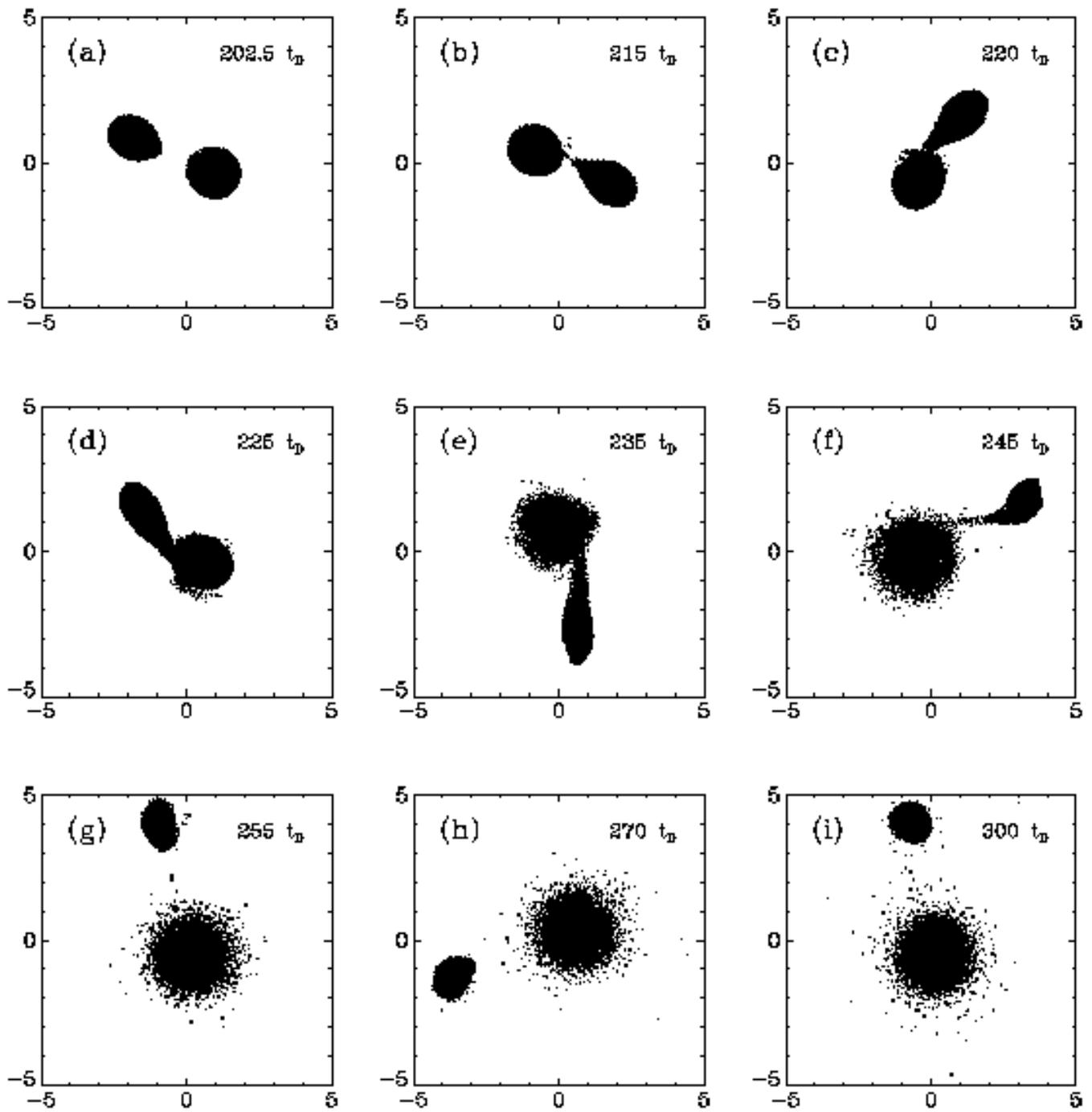
\begin{figure}[p]
\caption{Particle positions are shown projected onto the $x-y$
plane for Run 10.  Here, $M_1 = 1.4 {\rm M}_{\odot}$, 
$M_2 = 0.5 M_1$, $R_1 = 10$ km, $R_2 = 8.7$ km,  $\Gamma = 3$,
and $t_{\rm D} = 0.073$ ms.  The initial separation $a_0 = 4.0 R_1$.
The stars are orbiting in the counter-clockwise
direction.  The vertical axis in each frame is $y/R_1$ and the 
horizontal axis is $x/R_1$.  At the end of the run, frame (i),
the secondary has roughly 1/3 of its original mass.
\label{run10-snaps}}
\end{figure}

Run 10 uses $q = 0.5$ and $\Gamma = 3$, so
$R_2 = 8.7 {\rm km}$.  The evolution of this model is shown in
Figure~\ref{run10-snaps}.  As before, the tidal
 bulge on the secondary
grows and the resulting mass transfer starts to disrupt it.  The 
primary feels little effect as matter from the 
secondary spreads around
it.  However, this period of mass transfer ends 
$\lesssim 1.5$ orbits after it began when the 
secondary, now much reduced
in size, moves out to a wider orbit.  At the end of the simulation
($t = 300 t_{\rm D}$), the primary contains $\sim 86\%$ of the
total system mass and has a radius $\sim 1.5 R_1$, where $R_1$ is
the initial primary radius.  The secondary has a mass 
$\sim 0.29 {\rm M}_{\odot}$, about $42 \%$ of its original mass,
and is in orbit about the primary at
center of mass separation roughly $\sim 4.5 R_1$. 
Although we ended the
simulation at this point, in reality the inspiral will begin again
due to gravitational radiation reaction.

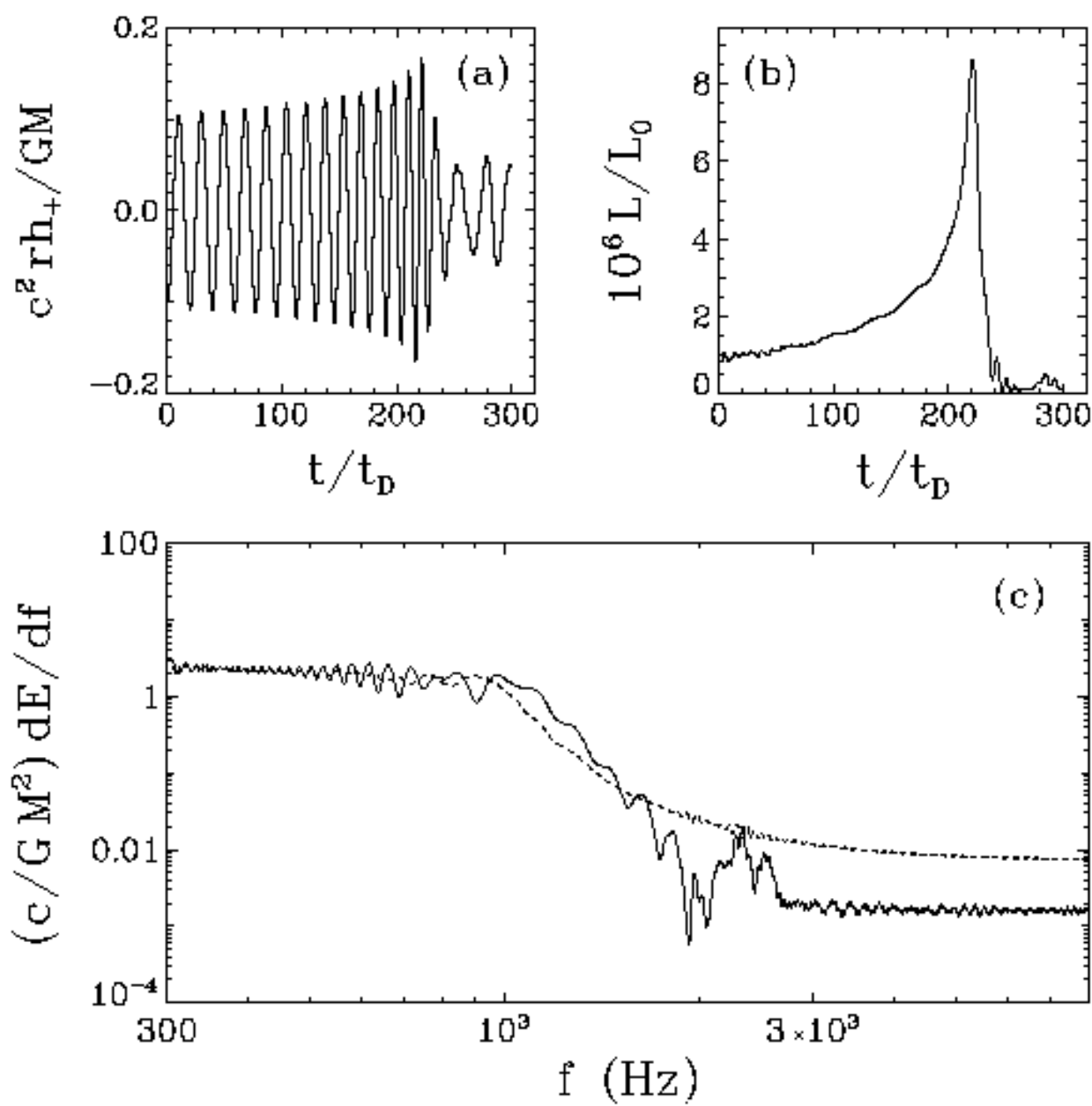
\begin{figure}[p]
\caption{
(a) The gravitational wave form $r h_+$ for Run 10 is shown for
an observer located on the axis at $\theta = \phi = 0$ at distance
$r$ from the source.
(b) The gravitational wave luminosity $L/L_0$,
where $L_0 = c^5/G$.
(c) The gravitational wave energy spectrum $dE/df$. The solid line
shows the spectrum for the entire run, and the 
dashed line shows the
spectrum for the wave forms truncated at $t = 220 t_{\rm D}$.
\label{run10-gw}}
\end{figure}

The gravitational wave form $r h_+$ is shown for this run in
Figure~\ref{run10-gw}(a), and the luminosity in
 Figure~\ref{run10-gw}(b).
Both the wave form and luminosity attain their 
maximum amplitudes in 
the early stages of mass transfer at 
$t \sim 220 t_{\rm D}$.  At late
times, the wave form shows the signature of the final binary orbit.
The maximum value of the wave form (see Table~\ref{models-max})
for a source located at distance r from the observer
is $(c^2/GM)r|h_{\rm max}| \sim 0.76(GM_1/R_1c^2)$ and
the peak luminosity is
$(L_{\rm max}/L_0) \sim 0.02(GM_1/R_1c^2)^5$.

The gravitational wave spectrum $dE/df$ for Run 10 is given in 
Figure~\ref{run10-gw}(c).  We find
$f_{\rm peak} \sim 900 {\rm Hz}$. The spectrum for the 
full run shows a signal in the region around
$\sim 600 {\rm Hz}$.  Since the point-mass gravitational wave
frequency for the end state of the system at separation 
$a \sim 4.5 R_1$ is $\sim 610 {\rm Hz}$,
we believe that this feature is due to the
final binary orbit.  In addition, a high-frequency feature 
$f_{\rm sec} \sim 2300 {\rm Hz}$ appears at late times.  Since the
amplitude of this secondary feature is so much smaller than the
main signal around $f_{\rm peak}$, we suspect that it may be 
a numerical artifact.

\section{Summary and Discussion}
\label{summary}

We have used SPH to perform 3-D simulations of the coalescence of
binary neutron stars with the goal of determining the gravitational
radiation signals produced and understanding the information that
can be extracted from the wave forms, luminosities, and spectra. 
The stars are initially modeled as spherical polytropes with masses
$M_1$ and $M_2$, radii $R_1$ and $R_2$, spins
$\Omega_{\rm s,1}$ and $\Omega_{\rm s,2}$, and equation of state
$\Gamma$, as summarized in Table~\ref{models-param}.  
At the start of each run, the stars are placed on 
(nearly) circular orbits with wide separations so that the binaries
are effectively in the point-mass limit.  The gravitational field
is purely Newtonian, and the gravitational radiation quantities
are calculated using the quadrupole approximation.  Frictional
terms are added to the equations of motion to mimic the effects of 
gravitational radiation reaction and to cause the stars to spiral 
together.  As the stars get closer, 
tidal effects grow and eventually
dominate.  At this point, the frictional
 terms are turned off and the
coalescence proceeds by purely Newtonian hydrodynamics.

Our first set of runs features binaries having components
with equal masses 
$M_1 = M_2 \equiv M = 1.4 {\rm M}_{\odot}$ and radii
$R_1 = R_2 \equiv R$.  We varied the radius $R$, equation of 
state $\Gamma$, and spin $\Omega_{\rm s}$.  In all of these
runs, coalescence occurs rapidly once the dynamical 
stability limit
is reached.  The merging stars form a rotating
barlike structure, and spiral arms are produced as mass is lost
 from the ends of the bar.  Gravitational torques cause angular 
momentum to be transported outward and lost to the spiral arms.  
The arms expand supersonically and merge to form a disk around the
central object.  For the cases that
 we studied, the rotating core of
the final merged remnant is axisymmetric for $\Gamma = 5/3$ and
$\Gamma = 2$, and non-axisymmetric for $\Gamma = 3$.  

The gravitational wave signals for these runs start out following
the point-mass results.  As the tidal effects grow stronger 
and the stars begin to spiral in faster than they 
would on point-mass trajectories, the amplitudes
and frequencies of the wave forms and the amplitude of the
luminosities increase faster than the point-mass results.
These amplitudes reach their maximum values in the early stages of
the coalescence, soon after the stars come into contact.  
Table~\ref{models-max} shows the scaling relationships
for these maximum values for our runs as well as four runs for
synchronous binaries by Rasio and Shapiro \cite{RS94} 
that are labeled RSa, b, c, and d.  
For non-spinning stars with $\Gamma = 2$, we find that
increasing the value of $\Gamma$ decreases the strength of
these maxima; cf. RSa and RSb.  
Allowing the stars to have identical spins
in the same direction as the orbital angular momentum also
increases the maximum values.  In particular, 
our Run 5 gives very similar results to RSa.
Interestingly, our Run 6, in which the stars have spins that
are equal in magnitude but opposite in direction, produces 
maximum values that are essentially the same as Run 1 with 
non-spinning stars.

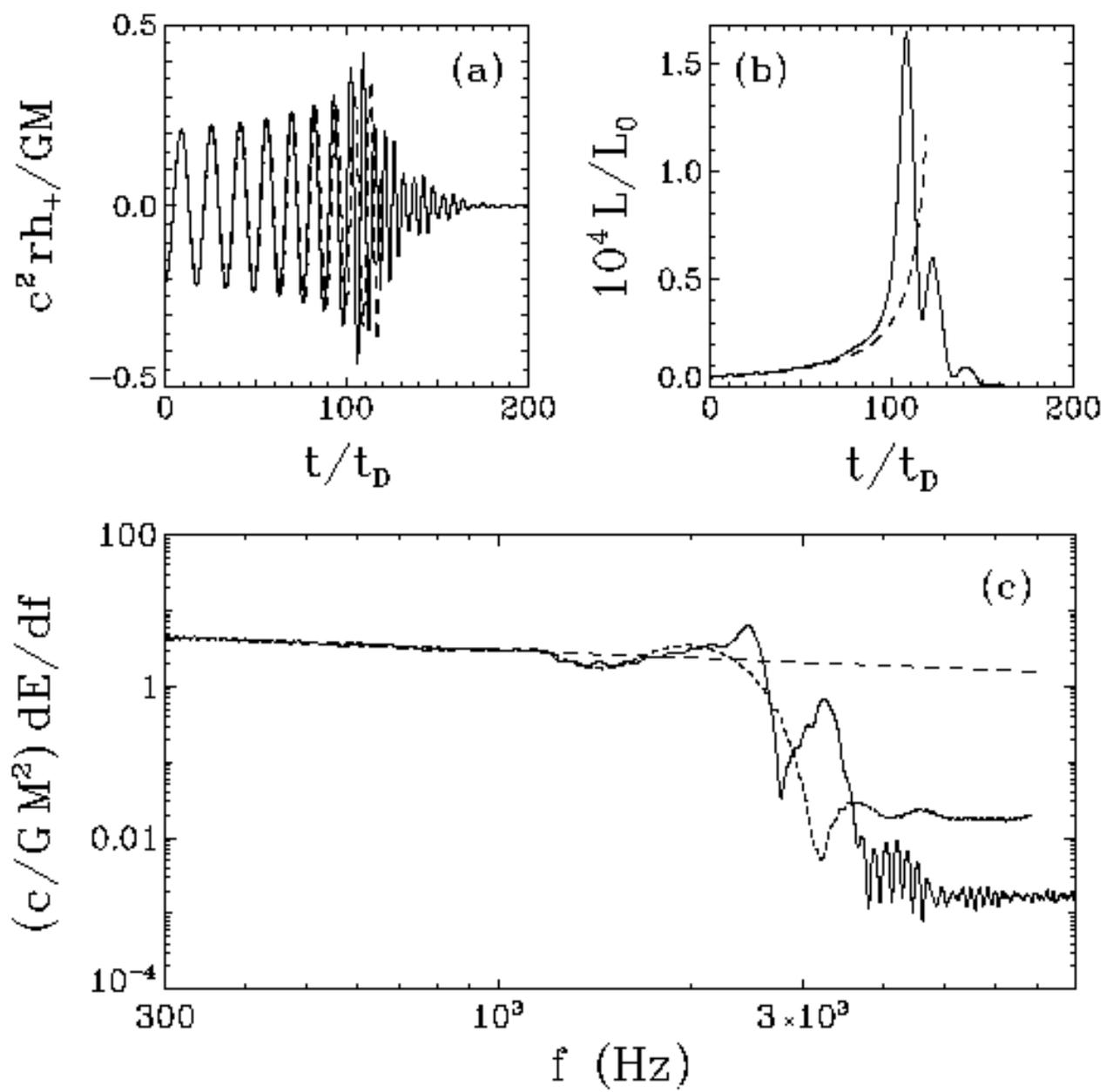
\begin{figure}[p]
\caption{
(a) The gravitational wave form $r h_+$ for Run 1 is shown for
an observer located on the axis at $\theta = \phi = 0$ at distance
$r$ from the source.  The solid line  is the code wave form and the
dashed line is the point-mass result.
(b) The gravitational wave luminosity $L/L_0$,
where $L_0 = c^5/G$. The solid line is the code result, and the
dashed line gives the point mass profile.
(c) The gravitational wave energy spectrum $dE/df$.
The code wave forms have been matched onto point-mass
results to produce a long region of point-mass inspiral
 at low frequencies.  The solid line
shows the spectrum for the entire run, and the 
short dashed line shows the
spectrum for the wave forms truncated at
 $t = 120 t_{\rm D}$. The long dashed is the point mass spectrum 
$dE/df \sim f^{- 1/3}$.
\label{run1-gw}}
\end{figure}

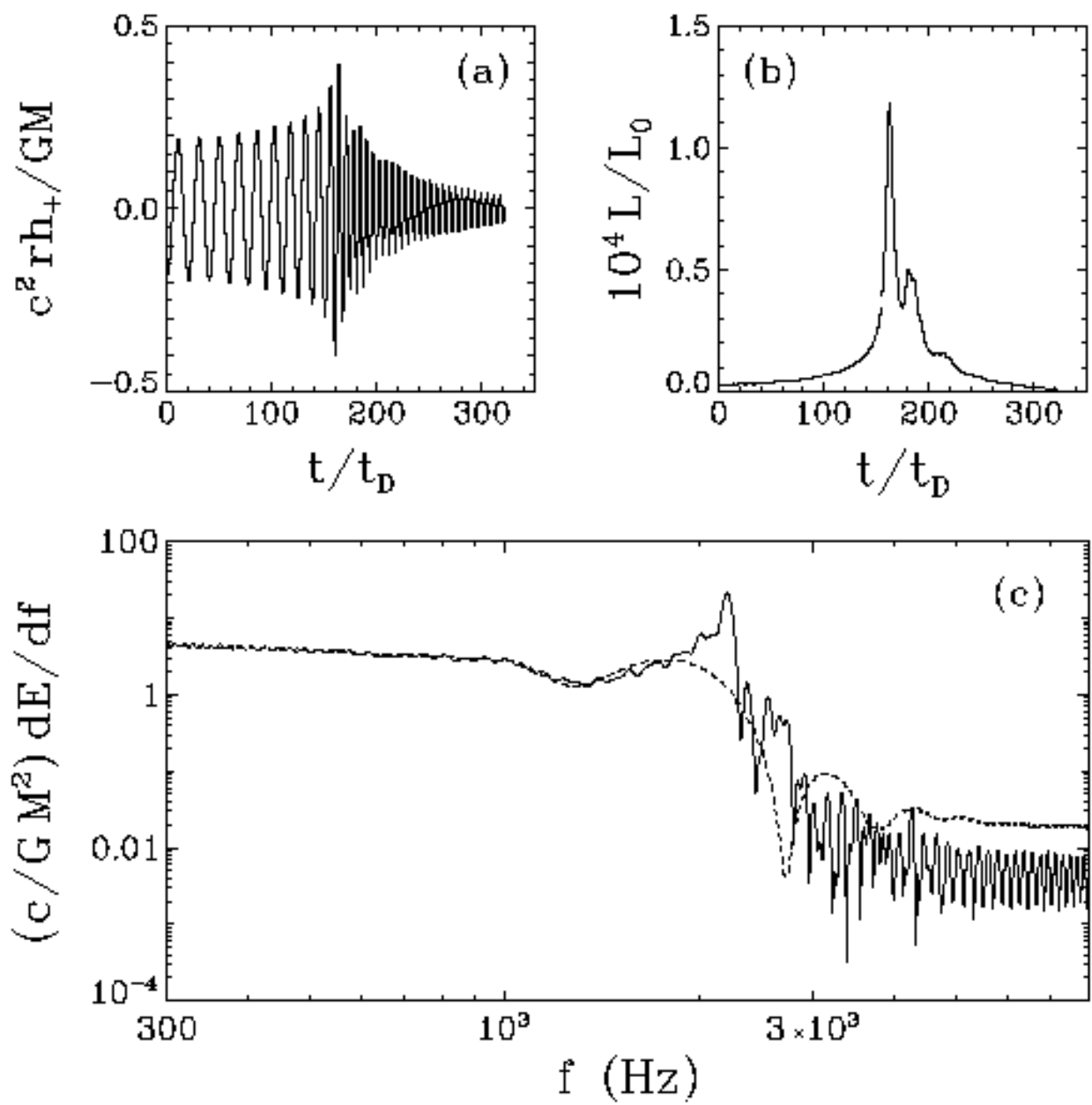
\begin{figure}[p]
\caption{
(a) The gravitational wave form $r h_+$ for Run 3 is shown for
an observer located on the axis at $\theta = \phi = 0$ at distance
$r$ from the source. 
(b) The gravitational wave luminosity $L/L_0$,
where $L_0 = c^5/G$.
(c) The gravitational wave energy spectrum $dE/df$. The solid line
shows the spectrum for the entire run, and the
 dashed line shows the
spectrum for the wave forms truncated at $t = 175 t_{\rm D}$.
\label{run3-gw}}
\end{figure}

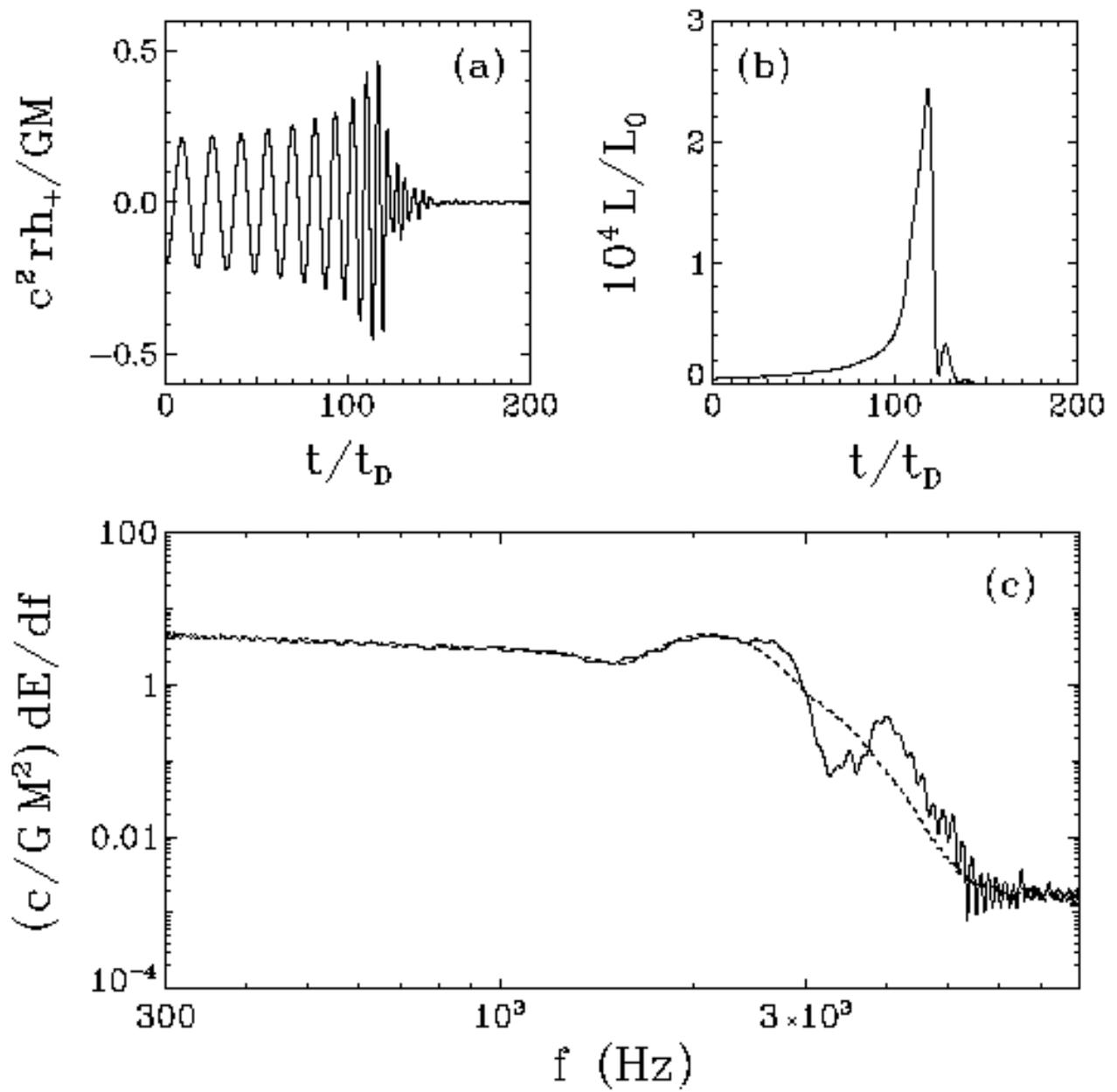
\begin{figure}[p]
\caption{
(a) The gravitational wave form $r h_+$ for Run 4 is shown for
an observer located on the axis at $\theta = \phi = 0$ at distance
$r$ from the source.
(b) The gravitational wave luminosity $L/L_0$,
where $L_0 = c^5/G$.
(c) The gravitational wave energy spectrum $dE/df$.
 The solid line
shows the spectrum for the entire run, and the
 dashed line shows the
spectrum for the wave forms truncated at $t = 125 t_{\rm D}$.
\label{run4-gw}}
\end{figure}

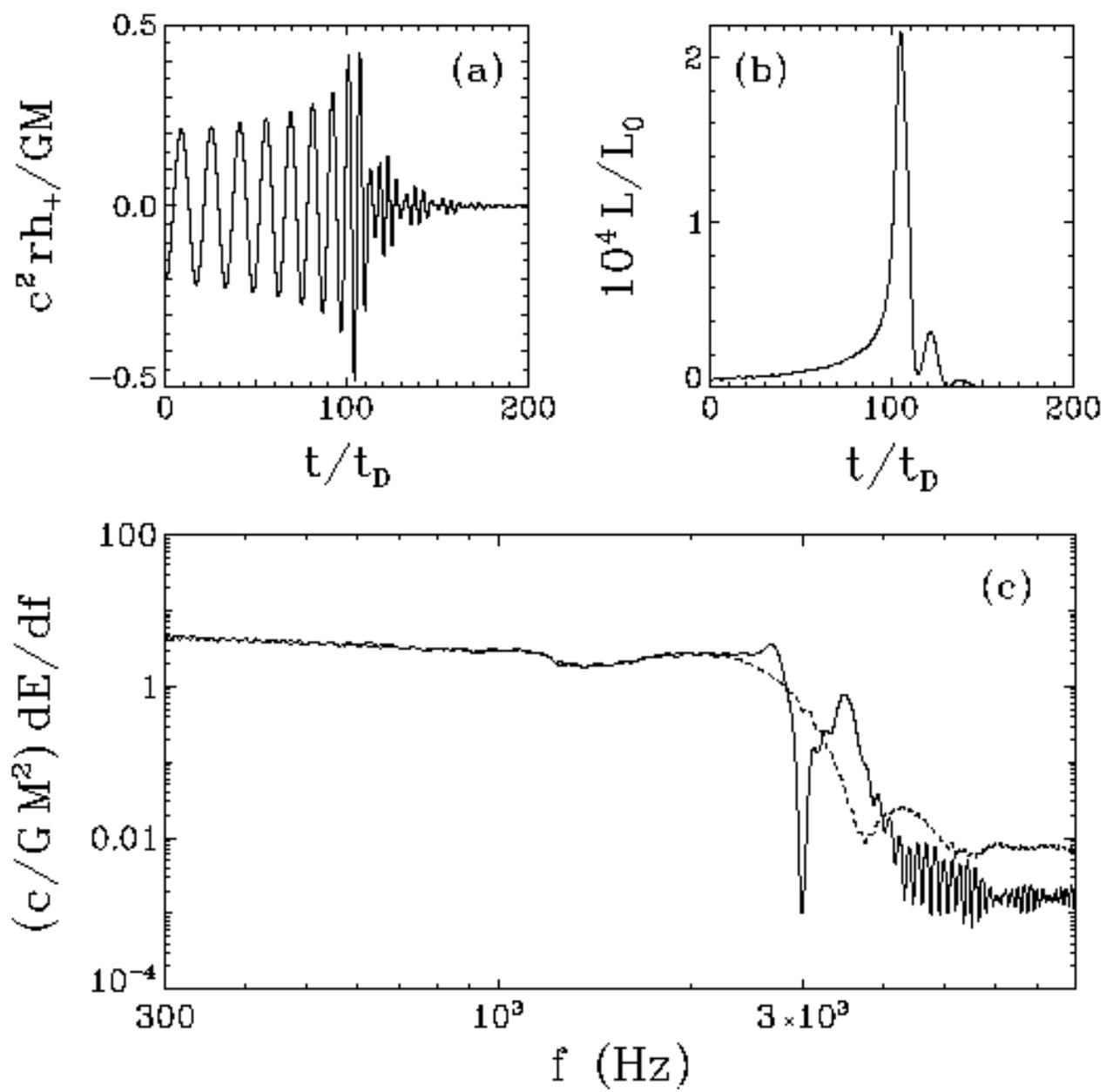
\begin{figure}[p]
\caption{
(a) The gravitational wave form $r h_+$ for Run 5 is shown for
an observer located on the axis at $\theta = \phi = 0$ at distance
$r$ from the source.
(b) The gravitational wave luminosity $L/L_0$,
where $L_0 = c^5/G$.
(c) The gravitational wave energy spectrum $dE/df$. The solid line
shows the spectrum for the entire run, and 
the dashed line shows the
spectrum for the wave forms truncated at $t = 112.5 t_{\rm D}$.
\label{run5-gw}}
\end{figure}

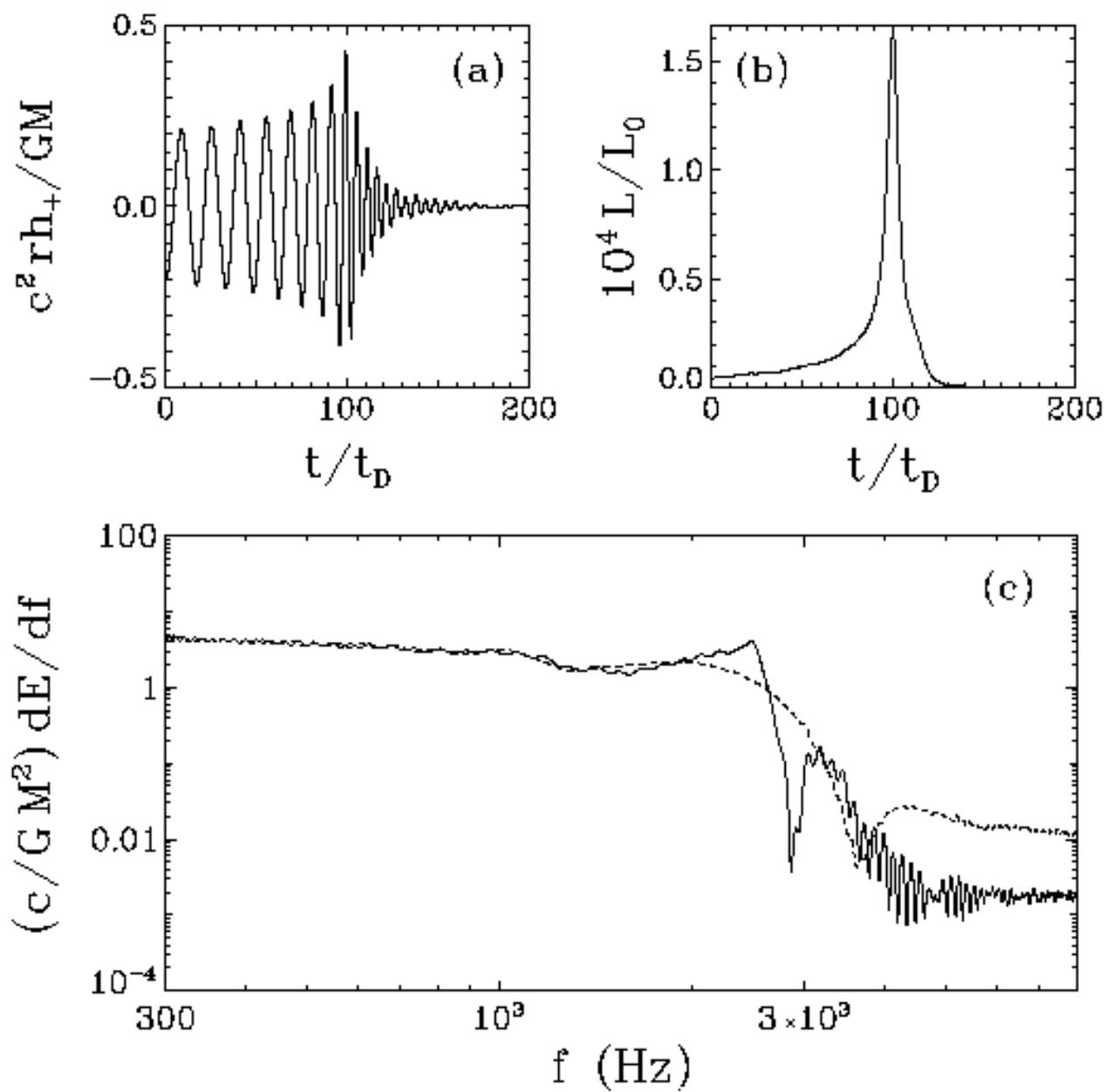
\begin{figure}[p]
\caption{
(a) The gravitational wave form $r h_+$ for Run 6 is shown for
an observer located on the axis at $\theta = \phi = 0$ at distance
$r$ from the source.
(b) The gravitational wave luminosity $L/L_0$,
where $L_0 = c^5/G$.
(c) The gravitational wave energy spectrum $dE/df$. The solid line
shows the spectrum for the entire run, and the dashed
 line shows the
spectrum for the wave forms truncated at $t = 107.5 t_{\rm D}$.
\label{run6-gw}}
\end{figure}

Examination of the wave forms and luminosities after the
maximum values are attained also reveals certain trends.
First consider the effects of changing $\Gamma$, starting with
Run 1 which has $\Gamma = 2$.  When
we decrease this parameter to $\Gamma = 5/3$ in Run 4, the waves
shut off more abruptly than in Run 1, as can be seen by examining
Figures~\ref{run1-gw}(a) and~\ref{run4-gw}(a).
We believe that this is due to the tendency for more compressible
fluids (i.e., those with smaller values of $\Gamma$) to reach the 
mass-shedding limit at smaller values of $\beta = T_{\rm rot}/|W|$.
However, when this parameter is increased to
$\Gamma = 3$ in Run 3 the wave form amplitude falls off more
gradually at late times due to the rotating,
 slightly non-axisymmetric
core; cf. Figures~\ref{run1-gw}(a) 
and~\ref{run3-gw}(a).  We attribute
this to the ability of polytropes with
 $\Gamma \gtrsim 2.4$ to sustain
ellipsoidal shapes since the mass-shedding 
limit along a sequence of
equilibrium models is reached after the point
 at which the ellipsoidal
sequence bifurcates from the spheroidal sequence. 
 When $\Gamma = 2$
and the stars have parallel spins in the direction of the orbital
angular momentum as in Run 5, the amplitude of the wave form drops
quite abruptly after the maximum is reached; this is shown in 
Figure~\ref{run5-gw}(a).  Similar behavior was seen by Rasio and
Shapiro \cite{RS92} and Shibata, et al. \cite{SNO}.  In Run 6
the spins are anti-parallel, and Figure~\ref{run6-gw}(a) shows that
the fall-off in the wave form is less abrupt than in the case of
parallel spins, but more rapid than in the case of zero spin.
In addition, the luminosity drops sharply after reaching its
maximum value.  All of these runs show 
three peaks with successively
decreasing amplitudes in the luminosity except for Run 6, which
has only one peak.  The synchronous runs of Rasio and Shapiro
\cite{RS92,RS94} show a single peak in the luminosity for 
$\Gamma = 5/3, 2$, and 3, with a secondary peak appearing for
$\Gamma = 10$.

The gravitational wave energy spectrum $dE/df$ is a
 useful tool for
analyzing the models.  In the point-mass regime, the spectrum
falls off according to $dE/df \sim f^{-1/3}$ \cite{thorne87}.
When tidal effects become dominant and the dynamical stability 
limit
is reached, the spectrum drops below the point mass curve and
reaches a local minimum around the corresponding frequency.  For
the early stages of the merger, the spectrum then rises to a
broad local maximum, then falls
off rather quickly at higher frequencies.  At later times, the peak
sharpens and moves to higher frequencies, due
to a transient, rotating bar-like structure that forms during the
coalescence.
The spectrum next drops off very sharply, and then rises to a 
secondary maximum.  While we can explain 
this secondary peak in terms
of oscillations in the final remnant, 
it is unclear how reliable the simulation is 
at late times.  Table~\ref{models-freq} 
shows that all runs with $R = 10 {\rm km}$ and $\Gamma = 2$
give very similar values for $f_{\rm peak}$.
When $R = 15 {\rm km}$, these features occur at lower frequencies
which scale as expected, roughly as $\sim R^{-3/2}$.
Changing the equation of
state to $\Gamma = 3$ and using $R = 10 {\rm km}$ produces a
somewhat smaller decrease in these frequencies, which we attribute
to the occurrence of the dynamical instability at a slightly larger
orbital separation.

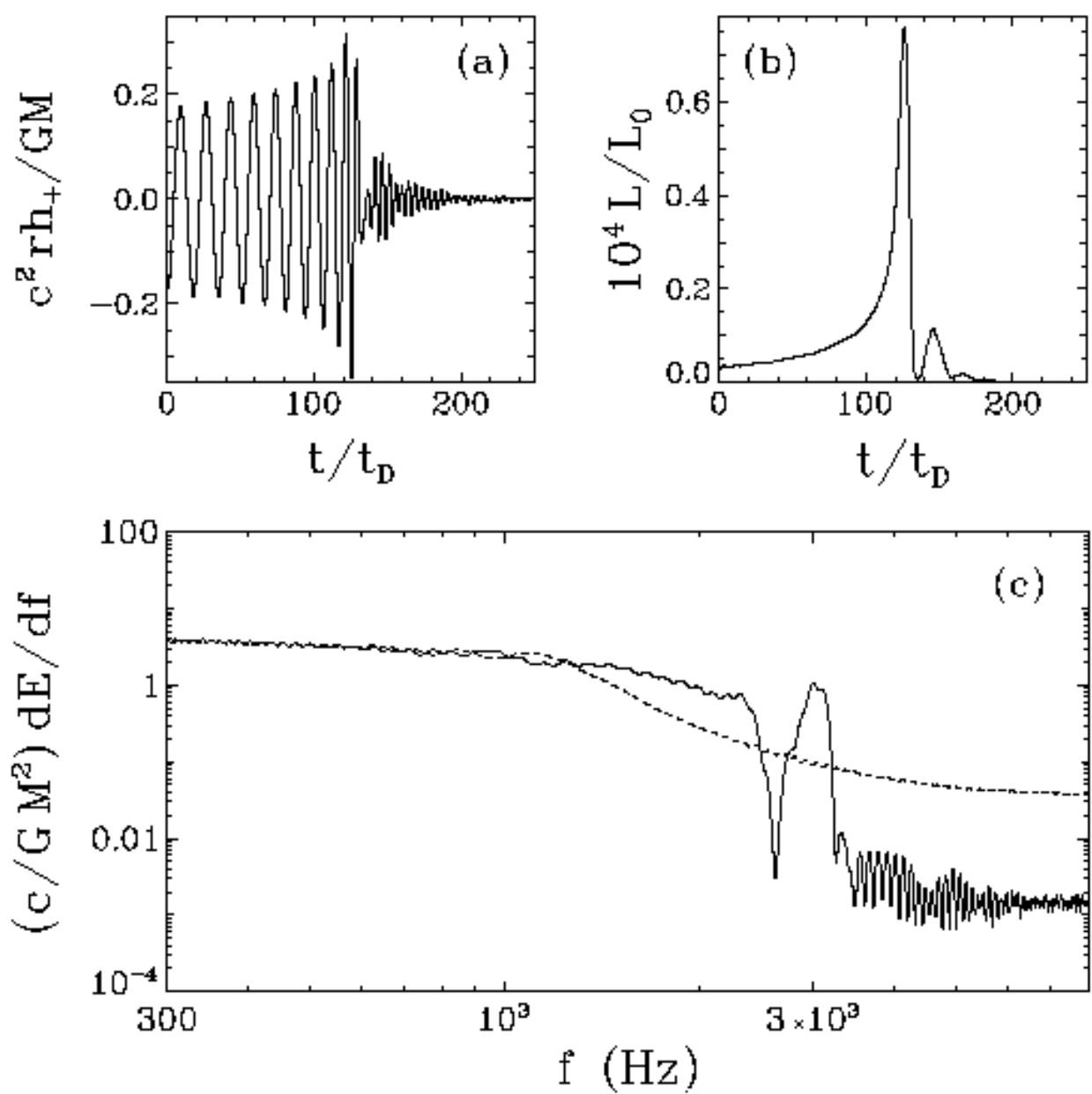
\begin{figure}[p]
\caption{
(a) The gravitational wave form $r h_+$ for Run 7 is shown for
an observer located on the axis at $\theta = \phi = 0$ at distance
$r$ from the source.
(b) The gravitational wave luminosity $L/L_0$,
where $L_0 = c^5/G$.
(c) The gravitational wave energy spectrum $dE/df$. The solid line
shows the spectrum for the entire run, and
 the dashed line shows the
spectrum for the wave forms truncated at $t = 125 t_{\rm D}$.
\label{run7-gw}}
\end{figure}

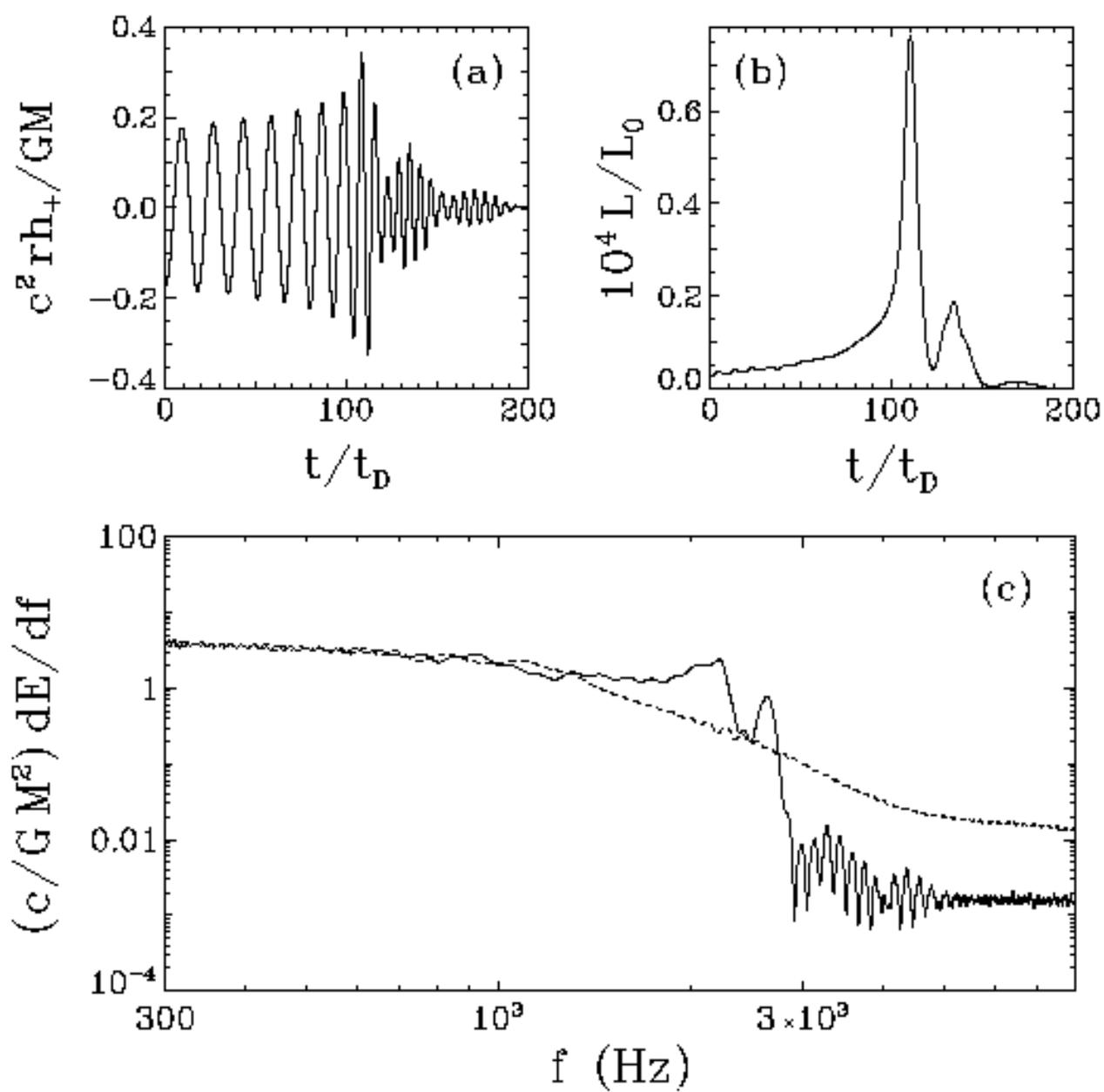
\begin{figure}[p]
\caption{
(a) The gravitational wave form $r h_+$ for Run 8 is shown for
an observer located on the axis at $\theta = \phi = 0$ at distance
$r$ from the source.
(b) The gravitational wave luminosity $L/L_0$,
where $L_0 = c^5/G$.
(c) The gravitational wave energy spectrum $dE/df$. The solid line
shows the spectrum for the entire run, and 
the dashed line shows the
spectrum for the wave forms truncated at $t = 112.5 t_{\rm D}$.
\label{run8-gw}}
\end{figure}

We performed two simulations with non-equal
 mass stars having mass
ratio $q = 0.85$.  In Run 7 with $\Gamma = 2$,
 the two stars merge to
form a remnant that has a central core of radius
 $\sim 1.5 R_1$ 
surrounded by a flattened halo formed out of matter
 from the secondary.
The behavior of Run 8 with $\Gamma = 3$ is similar,
 except that the
central rotating object remains elongated for a
 longer period of time.
This is reflected in the gravitational wave 
spectrum $dE/df$ shown 
as the solid line in Figure~\ref{run8-gw}(c),
 which shows a pronounced
peak at $f_{\rm peak} \sim 2200 {\rm Hz}$. 
In contrast, the spectrum for
Run 7, which is shown as the solid line in 
Figure~\ref{run7-gw}(c), has a much weaker feature at
$f_{\rm peak} \sim 2300 {\rm Hz}$.

We have also carried out two runs with non-equal
 mass components having
mass ratio $q = 0.5$.  In Run 9 with $\Gamma = 2$,
 the merger proceeds
by mass transfer that completely disrupts the secondary.  The 
gravitational waves shut off less than 1.5 orbits after the 
maximum amplitude is reached.  When $\Gamma = 3$, as in Run 10, 
mass transfer also initially disrupts the secondary.  However,
less than 1.5 orbits later this process stops as the secondary
moves out to a wider orbit.  At the end of the run the
 secondary has
$\sim 42\%$
of its original mass and orbits the primary at center of
mass separation $\sim 4.5 R_1$, where $R_1$ is the initial
radius of the primary.  Further evolution of this
system will proceed on a secular timescale by gravitational 
radiation reaction.  Rasio and Shapiro \cite{RS94} also report
the formation of a detached binary for the case of an initially
synchronous system with $\Gamma = 3$ and $q = 0.5$. 
Interestingly, their
maximum amplitudes for the wave form and luminosity are very
similar to ours (although their masses and final orbital 
separation are not); 
cf. Runs 10 and RSd in Table~\ref{models-max}.

It is important to understand the influence of numerical effects on 
the results of our simulations.  In Paper I we investigated the use
of various artificial viscosity coefficients.  We also showed that
our results do not depend strongly on $N$, the number of particles
per star, for $N$ = 1024, 2048 and 4096.  However, as
discussed in \S~\ref{coal-std} above, we suspect that numerical
effects due to Kelvin-Helmholtz instabilities may be influencing
the behavior of the shear layer that forms where the two merging
stars meet.  In particular, the bar phase of the evolution may be
artificially shortened and the final remnant may have different
properties.  These numerical effects should become severe 
{\em after} the maximum gravitational radiation amplitudes are
reached.  Thus, we believe that the scaling of 
the maximum amplitudes,
the shape of the spectrum $dE/df$ for the early stages of the 
merger, and the approximate location of the frequency 
$f_{\rm peak}$ are reliable.  However, the shape of 
the gravitational
radiation signals after the maximum values are reached, 
the {\em strength}
of the spectral feature at $f_{\rm peak}$, and the secondary
feature at $f_{\rm sec}$ may be affected 
somewhat by numerical processes, and we advise caution when
interpreting these results. We remark that 
the simulations of non-synchronous binaries carried
out by other groups are also expected to suffer from these 
problems; see \S~\ref{coal-std}.  More work is needed to clarify
these issues, including simulations with higher resolution.

The gravitational waveforms and spectra resulting from these
Newtonian simulations contain much information about the
hydrodynamics of the merger.  Of course, general
relativity is likely to bring in other physical effects that need
to be studied in order to understand the data expected
 from the detectors.  For example, Lai and Wiseman \cite{LW96}
have recently shown that the inclusion of 
certain general relativistic
effects along with Newtonian tidal processes
 causes the neutron stars
to begin their final plunge towards merger at larger orbital
separations, and hence at lower frequencies, than in the purely
Newtonian case.  Such information is potentially very 
important for the detection
of the gravitational wave signals from binary neutron star 
coalescence by LIGO and other detectors.
We intend to incorporate full general relativity in our 
future work.

\acknowledgments
  We thank N. Andersson,
J. Houser, D. Lai, R. Price, 
F. Rasio, and K. Thorne for interesting and helpful
communications, and L. Hernquist for supplying a copy of TREESPH.
We are grateful to an anonymous referee whose constructive
criticism helped improve this paper.
We thank J. Houser for supplying the initial 
rotating stars used in Runs 5 and 6, and 
acknowledge the assistance of J. Gao and J. Houser
in producing the figures.
This work was supported in part by NSF grants PHY-9208914 and
AST-9308005, and by NASA grant NAGW-2559.
  The numerical simulations
were run at the Pittsburgh Supercomputing Center
under grant PHY910018P.

\newpage 

\begin{table}[p]
\begin{center}
\begin{tabular}{ccccccccccc}
Model  & $q$ & $R_1$ &
 $R_2$ & $a_0$  & $t_{\rm D}$ & $T_{\rm s,1}$ &
$T_{\rm s,2}$ & $n$ & $\Gamma$ & $N$ \\ 
   &  &  (km) &  (km) &  (km) & (ms)
 & (ms) & (ms)  &  &  & \\
\tableline
Run 1 & 1   & 10 & 10  & 40 & 0.073 & 0   
& 0      & 1   & 2   & 4096 \\
Run 2 & 1   & 15 & 15  & 60 & 0.13  & 0   
& 0      & 1   & 2   & 1024 \\
Run 3 & 1   & 10 & 10  & 45 & 0.073 & 0   
& 0      & 1/2 & 3   & 4096 \\
Run 4 & 1   & 10 & 10  & 40 & 0.073 & 0   
& 0      & 3/2 & 5/3 & 2048 \\
Run 5 & 1   & 10 & 10  & 40 & 0.073 & 2.6 
& 2.6    & 1   & 2   & 4055 \\
Run 6 & 1   & 10 & 10  & 40 & 0.073 & 2.6 
& $-2.6$ & 1   & 2   & 4055 \\
Run 7 & 0.85 & 10 & 10  & 40 & 0.073 & 0   
& 0      & 1   & 2   & 4096 \\
Run 8 & 0.85 & 10 & 9.7 & 40 & 0.073 & 0   
& 0      & 1/2 & 3   & 4096 \\
Run 9 & 0.5 & 10 & 10  & 40 & 0.073 & 0   
& 0      & 1   & 2   & 4096 \\
Run 10 & 0.5 & 10 & 8.7 & 40 & 0.073 & 0   
& 0      & 1/2 & 3   & 4096 \\
\end{tabular}
\end{center}
\caption{Parameters of the models are given.  For all runs,
$M_1 = 1.4 {\rm M}_{\odot}$.  The mass ratio is $q = M_2/M_1$,
the stellar radii are $R_1$ and $R_2$, and the initial 
separation is $a_0$.
The dynamical time $t_{\rm D}$ 
is calculated using the parameters of star 1.
$T_{\rm s}$ gives the spin period of each star, with a
positive (or negative) value denoting a spin in the
same (or opposite) direction as the orbital angular
momentum.  The polytropic index $n$ and the $\Gamma$ 
refer to the equation of state.
Each star contains $N$ SPH particles.
\label{models-param}}
\end{table}

\begin{table}[p]
\begin{center}
\begin{tabular}{ccccc}
Model &
$(c^2/GM_1)r|h_{\rm max}|$ &
$(c^2/GM_1)r|h_{\rm max}|\alpha^{-1}$ &
$10^4 (L_{\rm max}/L_0)$ &
$(L_{\rm max}/L_0) \alpha^{-5}$ \\
\tableline
Run 1 & 0.43 & 2.0  & 1.6   & 0.39  \\
Run 2 & 0.29 & 2.1  & 0.21  & 0.39  \\
Run 3 & 0.40 & 1.9  & 1.2   & 0.29  \\
Run 4 & 0.48 & 2.3  & 2.4   & 0.59  \\
Run 5 & 0.48 & 2.3  & 2.2   & 0.59  \\
Run 6 & 0.42 & 2.0  & 1.6   & 0.39  \\
Run 7 & 0.33 & 1.6  & 0.75  &  0.18 \\
Run 8 & 0.33 & 1.6  & 0.76  &  0.19 \\
Run 9 & 0.14 & 0.67 & 0.044 & 0.011 \\
Run 10 & 0.16 & 0.76 & 0.083 & 0.020 \\
RSa   & ---  & 2.4  & ---   & 0.55  \\
RSb   & ---  & 2.2  & ---   & 0.37  \\
RSc   & ---  & 1.6  & ---   & 0.14  \\
RSd   & ---  & 0.8  & ---   & 0.018 \\
\end{tabular}
\end{center}
\caption{The maximum amplitudes of the gravitational wave form 
and luminosity are given for each model.  The value 
$(c^2/GM_1)r|h_{\rm max}| \sim 0.4$ corresponds to an amplitude
$h \sim 1.4 \times 10^{-21}$ for a source at distance 
$r = 20$Mpc (the approximate distance to the Virgo cluster of
galaxies). 
Scaled versions of these quantities, in terms of the parameter
$\alpha \equiv GM_1/R_1c^2$, are also presented.
The entries RSa,b, c, and d refer to models of synchronous binaries
run by Rasio and Shapiro and are taken from Table I in 
reference~\protect{\cite{RS94}}. Model RSa has $\Gamma = 2$ and
$q = 1$; compare this with our Runs 1, 2, 5 and 6.
Model RSb has $\Gamma = 3$ and $q = 1$; this should be
compared with our Run 3.   Model RSc has $\Gamma = 3$ and
$q = 0.85$ and
Model RSd has $\Gamma = 3$ and $q = 0.5$; compare these with our
Runs 8 and 10, respectively. 
\label{models-max}}
\end{table}

\begin{table}[p]
\begin{center}
\begin{tabular}{ccc}
Model & 
 $f_{\rm peak}$ (Hz) & $f_{\rm sec}$ (Hz) \\ 
\tableline
Run 1 & 2500 & 3200  \\
Run 2 & 1350 & 1750 \\
Run 3 & 2200 & 2600  \\
Run 4 & 2700 & 4000 \\
Run 5 & 2700 & 3500 \\
Run 6 & 2500 & 3200 \\
Run 7 & 2300 & 3000 \\
Run 8 & 2200 & 2600 \\
Run 9 &  940 & --- \\
Run 10 & 1000 & 2300 \\
\end{tabular}
\end{center}
\caption{Various frequencies of the models 
relating to the spectra $dE/df$ are given.
The frequencies $f_{\rm peak}$ and $f_{\rm sec}$
refer to the spectra calculated from the full wave forms and
and give the bar rotation speed and the oscillation frequency
for the remnant.  The truncated spectra peak at somewhat lower
frequencies.  (The value of $f_{\rm peak}$ given here for
Run 2 is somewhat smaller than that used in Paper I; the 
difference is due to some arbitrariness in estimating the
location of the ``cliff'' in Fig.~\protect{\ref{run2-gw}(c)}.)
\label{models-freq}}
\end{table}

\clearpage

\end{document}